\documentclass[trackchanges,twocolumn]{aastex}

\usepackage[utf8]{inputenc}
\usepackage{natbib,epsfig,siunitx,url,graphicx,amsmath,footnote,float,xcolor,cases}
\begin{document}

\submitted{Accepted for publication in Icarus}

\title{Mercury's formation within the Early Instability Scenario}

\author{Matthew S. Clement\altaffilmark{1}, John E. Chambers\altaffilmark{1}, Nathan A. Kaib\altaffilmark{2} Sean N. Raymond\altaffilmark{3} \& \& Alan P. Jackson\altaffilmark{4}}

\altaffiltext{1}{Earth and Planets Laboratory, Carnegie Institution for Science, 5241 Broad Branch Road, NW, Washington, DC 20015, USA}
\altaffiltext{2}{HL Dodge Department of Physics Astronomy, University of Oklahoma, Norman, OK 73019, USA}
\altaffiltext{3}{Laboratoire d'Astrophysique de Bordeaux, Univ. Bordeaux, CNRS, B18N, all{\'e} Geoffroy Saint-Hilaire, 33615 Pessac, France}
\altaffiltext{4}{School of Earth and Space Exploration, Arizona State University, 550 E Tyler Mall, Tempe, Arizona,85287, USA}
\altaffiltext{*}{corresponding author email: mclement@carnegiescience.edu}

\defcitealias{clement18}{Paper 1}
\defcitealias{clement18_frag}{Paper 2}
\defcitealias{clement21_tp}{Paper 3}
\defcitealias{clement20_psj}{C20}
\defcitealias{clement21_instb}{C21d}
\defcitealias{clement21_merc2}{C\&C21}
\defcitealias{clement21_merc3}{C21a}
\defcitealias{nesvorny12}{N\&M12}
\defcitealias{chambers98}{C\&W98}
\defcitealias{hansen09}{H09}

\begin{abstract}

The inner solar system's modern orbital architecture provides inferences into the epoch of terrestrial planet formation; a $\sim$100 Myr time period of planet growth via collisions with planetesimals and other proto-planets.  While classic numerical simulations of this scenario adequately reproduced the correct number of terrestrial worlds, their semi-major axes and approximate formation timescales, they struggled to replicate the Earth-Mars and Venus-Mercury mass ratios ($\sim$9 and 15, respectively).  In a series of past independent investigations, we demonstrated that Mars' mass is possibly the result of Jupiter and Saturn's early orbital evolution, while Mercury's diminutive size might be the consequence of a primordial mass deficit in the region (potentially the result of the growing Earth's early outward migration).  Here, we combine these ideas in a single modeled scenario designed to simultaneously reproduce the formation of all four terrestrial planets and the modern orbits of the giant planets in broad strokes.  By evaluating our Mercury analogs' core mass fractions, masses, and orbital offsets from Venus, we favor a scenario where Mercury forms through a series of violent erosive collisions between a number of $\sim$Mercury-mass embryos in the inner part of the terrestrial disk.  We also compare cases where the gas giants begin the simulation locked in a compact 3:2 resonant configuration to a more relaxed 2:1 orientation and find the former to be more successful.  In 2:1 cases, the entire Mercury-forming region is often depleted due to strong sweeping secular resonances that also tend to overly excite the orbits of Earth and Venus as they grow.  While our model is quite successful at replicating Mercury's massive core and dynamically isolated orbit, the planets' low mass remains extremely challenging to match.  Indeed, the majority of our Mercury analogs have masses that are 2-4 times that of the real planet.  Finally, we discuss the merits and drawbacks of alternative evolutionary scenarios and initial disk conditions (specifically a narrow annulus of material between 0.7-1.0 au).  We argue that the results of our N-body accretion models are not sufficient to break degeneracies between these different models, and implore future studies to apply further cosmochemical and dynamical constraints on terrestrial planet formation models.

\end{abstract}

\section{Introduction}
\label{sect:intro}

Models of the late stages of terrestrial planet formation \citep[e.g.:][]{wetherill80,ray18_rev} typically follow the dynamical evolution of a collection of $\sim$10-100 proto-planets \citep[dubbed ``embryos'', e.g.:][]{wetherill93,koko_ida_96,koko_ida00,morishma10} engulfed within a distribution of $\sim$100-1,000 km planetesimals \citep[e.g.:][]{youdin05,johansen15,draz18,lichtenberg21}.  As the giant planets' gaseous compositions indicate that they formed within the lifetime of the Sun's primordial gas disk \citep[$\lesssim$1-5 Myr:][]{haisch01,hernandez07}, their presence in simulations of the terrestrial world's ultimate accretion is essential \citep{chambers98,levison03}.  If the giant planets' orbits remain circular through the duration of terrestrial planet formation as predicted in hydrodynamical disk models \citep{papaloizou00,masset01,morbidelli07,pierens08,zhang10}, the final masses of Mars and the asteroid belt are too massive by at least an order of magnitude \citep{chambers01,ray06,ray09a,lykawka19,woo22}.  However, the moderate degree of radial mixing in such a scenario also has the advantage of aiding in the replication of Earth's water content \citep{raymond04} and disparities between the isotopic compositions of Earth and Mars \citep{tang14,dauphas17,woo21_tp}.

It is possible to generate a small Mars and low-mass asteroid belt with circular giant planet orbits if, rather than spanning the full radial range between Mercury and Jupiter's modern orbits, the majority of the terrestrial disk's mass is concentrated in a narrow annulus between the current orbits of Venus and Earth \citep{agnor99,hansen09,iz14}.  Multiple explanations (see section \ref{sect:models}) for these specific initial conditions have been proposed and robustly tested in the recent literature \citep[for a more complete summary see:][]{clement18,ray18_rev}.  Among others, these include material removal during the gas disk phase via sweeping secular resonances with eccentric giant planets \citep[the ``Dynamical Shake-up:''][]{nagasawa00,thommes08,bromley17,woo21}, Jupiter directly sculpting the terrestrial disk by migrating inward to $\sim$1.5 au and back out due to interactions with the nebular gas \citep[the ``Grand Tack:''][]{walsh11,pierens11,jacobson14,brasser16,deienno16,walsh16}, or highly localized planetesimal formation \citep[the ``low-mass asteroid belt'' or ``depleted disk'':][]{izidoro15,izidoro16,draz16,ray17sci,lykawka19,mah21,morby21,izidoro22_ss}.

An additional complication on this series of events is the giant planets' acquisition of their modern, moderately eccentric orbits \citep{hahn99,Tsi05} through an epoch of mutual encounters \citep{morby09,nesvorny11}.  While a low-mass Mars is a regular outcome in simulations where the giant planets' inhabit their current dynamical configuration for the duration of the simulation \citep{ray09a,kaibcowan15,lykawka19,woo21,nesvorny21_tp}, such a scenario conflicts with the predictions of many disk models of giant planet formation and early evolution \citep{morby07,pierens08,zhang10}, and also cannot explain Earth and Mars' disparate compositions \citep{woo22}.  

In the classic Nice Model \citep{gomes05,Tsi05,levison11}, the giant planets' orbits transition from circular to eccentric when they undergo an epoch of dynamical instability at $t\simeq$ 650 Myr; coincident with a perceived spike in lunar cratering known as the Late Heavy Bombardment \citep{tera74}.  However, simulations of a late instability typically result in the catastrophic disruption of the fully formed terrestrial system \citep{bras09,agnorlin12,bras13,kaibcham16}.  Moreover, multiple recently derived geophysical \citep[e.g.:][]{evans18,morb18,mojzsis19,brasser20}, geochemical \citep[e.g.:][]{boehnke16,zellner17,goodrich21,worsham21} and dynamical \citep[e.g.:][]{nesvorny18,quarles19,ribeiro20,nesvorny21_e_nep,liu22} constraints have been interpreted to strongly suggest that the instability happened within the first 100 Myr after the solar system's birth, and perhaps much earlier.

While an instability at $t\simeq$ 10-100 Myr might have destabilized a nearly-formed terrestrial system \citep{desouza21} and triggered the giant impact that formed the Moon \citep{benz86,canup04,canup12,cuk12} around the same epoch \citep[$\sim$30-100 Myr after nebular dispersal as inferred via isotopic dating:][]{wood05,kleine09,rudge10,kleine17}, an instability occurring before embryos at 1.5 au grow beyond a Mars-mass has been shown to reduce the final masses of Mars analogs and the asteroid belt \citep{clement18,deienno18,clement18_ab,clement21_tp,nesvorny21_tp}.  In this paper, we continue to develop this model with new numerical simulations.  In contrast to our past work (described below), our current effort specifically incorporates reduced integration timesteps and inner terrestrial disk structures necessary for generating Mercury-like planets \citep{clement21_merc2,clement21_merc3,clement21_merc4}.  While the distributions of planetesimals and embryos in the Mercury-region used here were found to be successful at producing reasonable Mercury analogs in a pervious series of studies (described below), it is currently unclear whether strong sweeping secular resonances with Jupiter in the inner solar system that occur during the giant planet instability would adversely affect Mercury's growth in such a scenario.  For this reason, the primary goal of this paper is to understand how the Nice Model instability might have affected the formation of Mercury. 

\section{Background}
\label{sect:past} 

\begin{figure*}
	\centering
	\includegraphics[width=1.\textwidth]{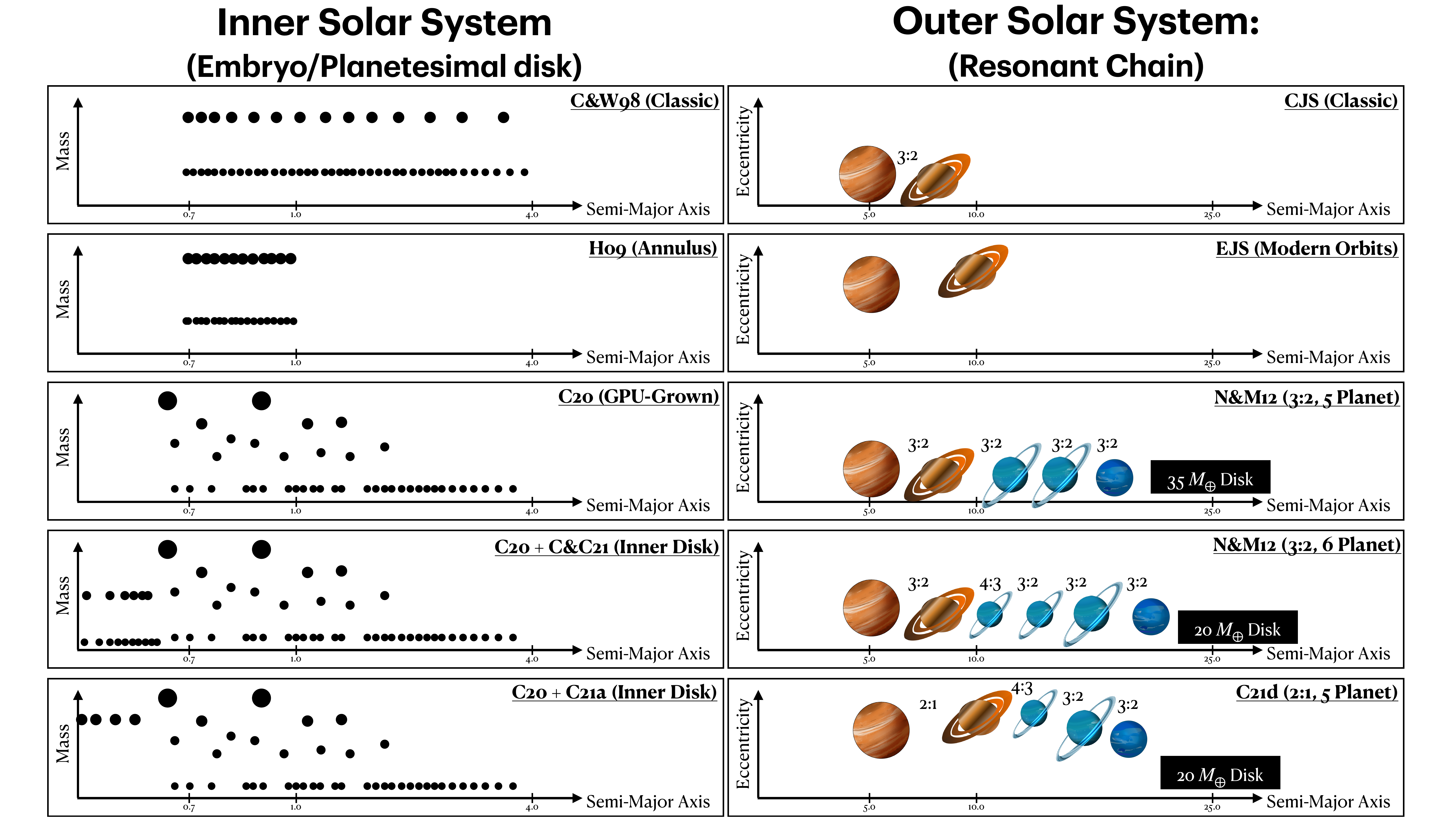}
	\caption{Cartoon representation of the various initial conditions tested here, and in our past investigations of the early instability scenario in \citetalias{clement18}, \citetalias{clement18_frag} and \citetalias{clement21_tp}.  The cartoon images of the outer planets are from www.kissclipart.com}
	\label{fig:cartoon}
\end{figure*}

 \subsection{Models replicating Mars' mass}
 \label{sect:models}
 \subsubsection{Giant Planet Migration: The Grand Tack}

Perhaps the most recognizable terrestrial planet formation scenario, the ``Grand Tack'' model \citep[as in the sailing maneuver:][]{walsh11,jacobson14,brasser16} supposes that the inward-outward migration \citep{pierens11} of Jupiter and Saturn during the gas disk phase truncated the terrestrial disk of planetesimals around $\sim$1.0 au \citep{wetherill78,morishima08,hansen09}.  Specifically, Jupiter's presence in the terrestrial region serves to both evacuate a large fraction of material from the Mars-forming region and simultaneously transfer volatile-rich C-type asteroids from the outer solar system to the asteroid belt \citep{demeo13}.  In spite of these consistencies, the mechanism for Jupiter's tack strongly depends on unconstrained disk parameters, and has yet to be validated in simulations incorporating gas accretion \citep{raymorb14}.  Additionally, the strong radial mixing of material that occurs in the scenario is potentially inconsistent \citep{mah21} with the disperate isotopic compositions of Earth \citep{lodders00,dauphas17} and Mars \citep{tang14}.
 
\subsubsection{Secular Resonance Sweeping: The Dynamical Shake-up}
 
If Jupiter's orbit was eccentric while gas persisted in the solar nebula, the powerful $\nu_{5}$ resonance would have swept from the belt's outer edge, inward to the vicinity Earth's orbit as the disk photo-evaporated \citep[the ``Dynamical Shake-up'' described in:][]{nagasawa00,thommes08,bromley17}.  While early hydrodynamical models \citep{masset01,morby07} indicated that such primordial eccentricities are unlikely outcomes of the gas disk phase, recent work suggests that primordial excitation is plausible for certain disk parameters \citep{kley04,zhang10,pierens14}.  Moreover, the giant planets' modern configuration \citep{clement21_instb} and the disparate accretion zones of Earth and Mars \citep{woo21_tp} are broadly consistent with such a scenario.
 
\subsubsection{Highly Localized Planetesimal Formation Efficiency: The Low-Mass Asteroid Belt}
 
Modern models of planetesimal formation have demonstrated how the process is highly sensitive to the local thermal and structural properties in the disk \citep{simon16,draz16,lichtenberg21}.  Moreover, ALMA observations showing non-uniform radial concentrations of dust in proto-planetary disks seem to support highly localized, or radially dependent planetesimal formation.  The low-mass asteroid belt model \citep{izidoro15,ray17,ray17sci,izidoro21_tp} proposes that very little solid material ever existed in the Mars-forming and primordial asteroid belt regions, and the terrestrial planets grew from a correspondingly steep surface density profile of material.   However, it remains difficult to determine how realistic such a distribution of solid material is.  In this manner, it is challenging for disk models to demonstrate the contemporary formation of two isotopically distinct populations of iron-meteorite parent body planetesimals at disparate radial locations \citep[see:][]{lichtenberg21,morby21,izidoro22_ss,chambers22}.

\subsubsection{Terrestrial migration: Convergent and outward}
 
 It is also possible that the terrestrial embryos themselves migrated from their initial formation locations.  Indeed, depending on the particular physical structure and thermal profile of the nebular disk's inner component, inward, $\gtrsim$Mars-mass embryos can experience inward, outward or convergent migration \citep[e.g.:][]{cresswell08,lyra10,paardekooper11,bitch15,eklund17}.  Recently, \citet{broz21} leveraged this concept to demonstrate that the masses and orbits of all four terrestrial planets might be a consequence of a more diffuse collection of terrestrial embryos ($\sim$0.4-1.8 au) being reshaped via convergent migration towards $\sim$1 au.  Similarly, \citet{clement21_merc4} invoked outward migration of Earth and Venus' precursor embryos to reconcile isotopic differences between the Earth and Mars \citep{tang14,dauphas17}, as well as provide and explanation for Mercury's diminutive mass and isolated orbit \citep{clement21_merc2}.  However, these models rely heavily on \textit{a priori} assumptions of the solar nebula's structure.

\subsection{The Early Instability Scenario}

In \citet[][hereafter Paper 1]{clement18} we studied the effects of the Nice Model instability on the forming terrestrial planets by varying the time at which the instability transpires within the accretion process.  Here, and throughout this manuscript, we refer to the ``instability delay'' as the amount of time a terrestrial planet formation simulation progresses before the instability ensues.  We loosely correlate the beginning of the accretion   simulations with nebular dissipation (or 2-3 Myr after the formation of Calcium Aluminum-Rich Inclusions: CAIs); however, this connection is not exact.  

In all instability models tested, the terrestrial planets formed from an extended disk of 100 embryos and 1,000 planetesimals with an outer boundary at 4.0 au \citep[modeled after the classic initial conditions of][hereafter C\&W98, see figure \ref{fig:cartoon}]{chambers98}, and the giant planet instability was modeled using the preferred initial conditions of \citet[][hereafter referred to as the N\&M12 model]{nesvorny12}.  Specifically, the \citetalias{nesvorny12} instability model assumes that Jupiter and Saturn emerged from the gas disk locked in a 3:2 mean motion resonance (MMR).  In \citetalias{clement18}, we considered any simulation where Jupiter and Saturn's final orbital period ratio ($P_{S}/P_{J}=$ 2.49 in the modern solar system) finished less than 2.8 to be successful.  We tested five and six planet giant planet models of this type (specifically, resonant chains of the form 3:2,3:2,3:2,3:2 and 3:2,4:3,3:2,3:2,3:2\footnote{While the original Nice Model \citep{Tsi05,gomes05,mor05} only considered the four known outer planets, recent modifications include an additional one or two primordial ice giants to increase the probability of a simulation finishing with the correct number of planets, and reduce the amount of time powerful secular resonances spend in the asteroid belt and inner solar system \citep{bras09} by forcing Jupiter and Saturn's semi-major axes to evolve in step-wise manner upon the ejection of the additional planet(s) \citep{nesvorny11}.}), and instability delays of 0.01, 0.1, 1.0 and 10.0 Myr.

The major conclusions of our initial paper were that the instability efficiently limits Mars' mass while essentially setting its geologic accretion timescale without disturbing Earth and Venus' formation, in reasonable agreement with geochemical studies arguing that the planet finished forming within the first few Myr after nebular dispersal \citep{Dauphas11,tang14,kruijer17_mars}.  Our most successful outcomes occurred in simulations that most closely matched Jupiter's modern eccentricity, and those where the instability was delayed 1-10 Myr.  With shorter delays, the truncated terrestrial disk tended to spread back out and produce under-mass Earth and Venus analogs.  However, none of our simulation sets provided good matches to the inner planets' low degree of orbital excitation \citep[an outstanding problem in most formation models:][]{ray09a,lykawka19} and Mercury's low mass \citep[although generating such planets likely requires modifications to the C\&W98 disk conditions:][see further discussion in $\S$ \ref{sect:merc_ics}]{chambers01,lykawka17}.

In a follow-on study \citep[][hereafter Paper 2]{clement18_frag}, we essentially reran the simulations from \citetalias{clement18} utilizing a code that accounts for imperfect accretion by generating collisional fragments \citep[][we use this same code in our current study, see $\S$ \ref{sect:sims}]{chambers13}.  We leveraged the same \citetalias{nesvorny12} instability models, however we deviated from the methodology of our initial paper by stopping and discarding simulations where Jupiter and Saturn exceeded $P_{S}/P_{J}=$ 2.8.  In addition to the classic \citetalias{chambers98} disk, we also tested a narrow annulus of embryos and planetesimals confined between 0.7 and 1.0 au \citep[][hereafter referred to as the H09 disk]{hansen09} in both instability and control (giant planets on static orbits) models.  The major conclusions of \citetalias{clement18_frag} were that collisional fragmentation aids in lowering the eccentricities and inclinations of Earth and Venus analogs, and that the \citetalias{hansen09} annulus is compatible with the early instability scenario.  However, while a few systems displayed levels of dynamical excitation comparable to that of the real terrestrial system, such outcomes were rare.  Similarly, while small Mercury analogs formed as the result of fragmenting collisions in some of our models \citep[e.g.:][]{asphaug14}, the orbits of these planets were systematically too close to Venus' \citep[see][]{clement19_merc}.

\subsection{Terrestrial disks derived from high-resolution embryo formation models}

Initial conditions similar to our \citetalias{chambers98} and \citetalias{hansen09} disk are typically justified in the literature by the results of dust evolution models \citep{birnstiel12} and semi-analytic studies of runaway planetesimal accretion \citep{wetherill89,koko_ida_96,koko_ida_98,chambers06}.  Our investigations in \citetalias{clement18} and \citetalias{clement18_frag} indicated that the instability's efficiency of limiting Mars' mass without disturbing the other planets' formation and orbits is related to the partitioning of mass between embryos and planetesimals in each region ($M_{Emb}/M_{Pln}$).  To better ascertain the authentic values of $M_{Emb}/M_{Pln}$ at various regions of the terrestrial disk around the time of nebular dissipation, we performed high-resolution planetesimal accretion simulations in \citet[][hereafter referred to as the C20 disk]{clement20_psj}.  Our N-body models began with $r\simeq$ 100 km planeteimsals, utilized a GPU-accelerated integration package \citep{genga} and included algorithms designed to mimic the effects of the decaying gas disk \citep{morishma10}.  The main findings of our study were that embryos in the Earth/Venus region grow to $\sim$0.3-0.4 $M_{\oplus}$, and most of the small planetesimals in the vicinity are accreted within the life of the gas disk.  Contrarily, embryos do not form in the asteroid belt given the slow accretion timescales.  These results are largely consistent with other recent high-resolution modeling efforts \citep{carter15,walsh19,woo21}, and we validated the compatibility of the inferred disk structures within the early instability scenario in \citet[][hereafter Paper 3]{clement21_tp}.  In that work we found that the dominant effect of the updated initial conditions was to shorten the planets' geologic growth timescales.  In this manuscript, we utilize initial conditions based off these \citetalias{clement20_psj} disks in all of our new simulations.

\subsection{Updated instability evolutions}

Our analyses in \citetalias{clement18} and \citetalias{clement18_frag} demonstrated how the reduction of Mars' mass is related to the proper excitation of Jupiter's eccentricity; in particular the magnitude of its' fifth eccentric mode $e_{55}$ \citep[the term related to the solar system's $g_{5}$ eigenfrequency:][]{nobili89,morby09}.  However, the adequate replication of this quality is a low-likelihood event in simulations of the \citetalias{nesvorny12} instability that do not over-excite Saturn's eccentricity \citep{nesvorny12,deienno17}.  While classic studies of the Nice Model relied on the findings of one-dimensional, fixed-viscosity hydrodynamical disk models that Jupiter and Saturn most likely to be captured in a 3:2 MMR \citep{masset01,morby07,pierens08}, recent work demonstrated a broader spectrum of potential evolutionary pathways \citep[including capture in the 2:1 with primordial eccentricity excitation:][]{pierens14}.  In \citet[][hereafter referred to as the C21d instability model]{clement21_instb} we performed a broad investigation of the untested parameter space applicable to the eccentric, primordial 2:1 Jupiter Saturn resonance and found many cases that greatly increase the probability of replicating many properties of the outer solar system (especially Jupiter and Saturn's precise eccentricities) when compared to the \citetalias{nesvorny12} model.  Thus, we concluded that the scenario represents a viable evolutionary path for the solar system.  Around half of the terrestrial planet formation simulations we performed in \citetalias{clement21_tp} used the \citetalias{clement21_instb} instability model.  While \citetalias{nesvorny12} also investigated the 2:1 Jupiter-Saturn resonance, they did not consider the possibility that the gas giants' primordial eccentricities were elevated.. While the differences between the 3:2 and 2:1 cases were mostly minor in terms of their effects on the inner solar system, we did note a boosted efficiency of Mercury-analog formation in our 2:1 cases.  Most of these analogs were embryos liberated from the terrestrial disk during the instability.

\subsection{Dynamical avenues for Mercury's origin}
\label{sect:merc_ics}

Similar to many notable statistical studies of terrestrial planet formation leveraging N-body simulations \citep[e.g.:][]{ray09a,jacobson14,izidoro15}, we largely neglected Mercury's formation in \citetalias{clement18}, \citetalias{clement18_frag} and \citetalias{clement21_tp}.  As discussed above, forming Mercury analogs with the correct mass and orbital offset from Venus requires the use of specific initial conditions in the inner disk \citep[$a\lesssim$ 0.7 au, e.g.:][]{chambers01,lykawka17,lykawka19}, exotic planetary migration schemes during the gas disk phase \citep[e.g.:][]{batygin15a,ray16,broz21} or low-probability collisional geometries \citep[e.g.:][]{benz07,asphaug14,jackson18,chau18}.  Studying Mercury's formation within the early instability scenario is the primary goal of this paper.

As a starting point for our current investigation, we use the results of a pair of recent companion investigations by our group designed to identify viable formation avenues for Mercury.  In \citet[][hereafter C21a]{clement21_merc3}  we found that well-spaced ($\sim$30-40 mutual Hill Radii: $R_{H}$) systems of 4-6 $\sim$Mars-mass short-period planets are easily destabilized in a manner that often leaves behind a relic Mercury analog with the appropriate mass, orbit and composition.  In \citet[][hereafter C\&C21]{clement21_merc2} we identified a narrow subset of mass-depleted inner disk parameters (in terms of total mass, $M_{Emb}/M_{Pln}$ and surface density profile) that boosts the in-situ production efficiency of Mercury-like planets.  While we did not initially provide a physical motivation for these structures, in subsequent work \citep{clement21_merc4} we argued that both could be the product of proto-Earth and Venus first forming closer to the Sun before migrating out near the end of the gas disk's life.

The disparate interpretations of various geochemical and observational constraints in the literature could be invoked to either support or conflict with our proposed Earth/Venus outward migration scenario \citep[for a complete discussion of these caveats we refer the reader to][]{clement21_merc4}.  As the precise structure of the primordial solar nebula remains rather unconstrained, it is our intention to avoid presenting a thorough justification of the likelihood of our chosen initial conditions occurrence within the larger context of the young solar system's global evolution.  Instead, our strategy here is to begin our study with the inner disk configurations that are seemingly most likely to produce adequate Mercury-analog planets.  In this manner, the question of whether these proposed scenarios can be disentangled from the processes of terrestrial planet formation and the giant planet instability remain largely unaddressed.  For instance, the \citetalias{clement21_merc3} models mostly considered fully-formed versions of Earth, Venus and Mars, and only presented a small number of simulations that incorporated a \citetalias{hansen09} annulus of embryos and planetesimals.  The subsequent sections present the results of new simulations of the early instability scenario that modify the \citetalias{clement20_psj} disks utilized in \citetalias{clement21_tp} with inner disk components derived from \citetalias{clement21_merc3} and \citetalias{clement21_merc2}, and test both the \citetalias{nesvorny12} and \citetalias{clement21_instb} instability models.

\section{Methods}

\subsection{New simulations}
\label{sect:sims}

We utilize the same general methodology for generating initial conditions, triggering the instability and selecting the evolutions that best replicate the modern outer solar system as described in \citetalias{clement21_tp}.  Simply put, our full ``instability plus terrestrial planet formation'' simulations are launched on a number of compute cores, and simulations are continuously restarted with a new set of initial conditions if our algorithm detects that the post-instability state of the outer solar system is unsatisfactory as determined by $P_{S}/P_{J}$ and $e_{J}$.  Specifically, we discard simulations that do not excite $e_{J}$ to greater than 0.03, and those that exceed $P_{S}/P_{J}=$ 2.5.  Similarly, we stop any simulation where an $e_{J}$ output is less than 0.01 (half of Jupiter's modern minimum eccentricity) at any point after the instability.  Integrations that obtain a successful instability are run up to $t=$ 200 Myr and saved for analysis.  We also perform an additional 10 Myr simulation of the final planetary system (without small bodies) using a higher coordinate output frequency to calculate the secular frequencies and magnitudes via frequency modulated Fourier transform \citep{nesvorny96}.

Discarding all simulations that exceed a Jupiter-Saturn period ratio of 2.5 severely limits our sample size of evolved systems when compared to our previous studies \citepalias{clement18,clement18_frag,clement21_tp} that included simulations that finished in the 2.5 $<P_{S}/P_{J}<$ 2.8 range.  Indeed, of the thousands of systems we initialized for this work, only around $\lesssim$10$\%$ adequately excite Jupiter's eccentricity without pushing Saturn past its 5:2 MMR with Jupiter in the immediate aftermath of the instability.   Nearly three quarters of that subset of successful runs subsequently exceed $P_{S}/P_{J}=$ 2.5 via residual migration or follow-on dynamical instabilities.  However, through this process we are still able to obtain a sample of approximately 20 fully evolved systems for each set of initial conditions tested.  This allows us to make comparisons between the distributions of outcomes in the different simulation batches with a modest degree of statistical significance.  For this reason though, it is important to acknowledge that the strongest conclusions from our modeling work are those derived from comparisons of the different disk (i.e.: grouping all \citetalias{hansen09} annulus runs and scrutinizing them against the results of all \citetalias{chambers98} disks) or those contrasting the disparate instability models. 

It is valid to criticize our methodology of combining the results from a suite of instability realizations (none of which exactly replicate the outer solar systems' true early evolution) as being idealized.  However, alternative methodologies that utilize interpolation algorithms to force the giant planets to follow ideal evolutions that are hand-selected from large suites of instability simulations \citep{nesvorny13,nesvorny14a,nesvorny14b,roig15,roig16,deienno16,nesvorny21_tp,desouza21} are plagued by the same inherent weakness.  It is impossible to know whether the selected instability simulation represents the true evolution of the giant planets or not.  In our approach \citepalias[here, as well as in][]{clement18,clement18_frag,clement21_tp}, we attempt to select a range of instabilities that most closely replicate the modern dynamical configuration of Jupiter and Saturn.  While Uranus and Neptune's dynamical perturbations in the inner solar system during its formation \citep{chambers01,levison03}, and today \citep{nobili89,laskar90}, are minor; excitation of planetesimals in the terrestrial disk is closely related to Jupiter's free eccentricity \citep{ray09a}, and the Jupiter-Saturn period ratio consequentially sets the location of secular resonances that scatter and remove material from the asteroid belt \citep{minton10,walshmorb11,clement20_mnras}.  By combining a number of different evolutions that replicate these qualities in broad strokes, our work aims to present a large sample of potential early evolutions of the young solar system; each of which possess a disparate sequence of ice giant encounters, peak gas giant eccentricities, ice giant low perihelia passages, Jovian semi-major axis jumps, and residual migration distances.  However, it will be important to study whether any of our most successful instabilities (in terms of inner solar system constraints) present systematic issues for small body populations in the outer solar system \citep[e.g.:][]{nesvorny13,nesvorny14a,deienno14,nesvorny15a} in future work.

Table \ref{table:ics2} summarizes the important assumptions of our different simulation sets, along with the parameters utilized in a number of simulation sets from our past early instability papers used here for the purpose of comparison.  All simulations leverage a version of the \textit{Mercury6} Hybrid integrator \citep{chambers99} that includes modifications \citep{chambers13} designed to model imperfect accretion and collisional fragmentation \citep{genda12,leinhardt12,stewart12}.  When the angle and velocity of an impact place it into the erosive regime, the fragmenting mass is divided into a number of equal-mass particles with $m\geq$ 0.005 $M_{\oplus}$ that are ejected at uniformly-spaced directions in the collisional plane with $v\simeq$ 1.05$v_{esc}$.  While the inclusion of collisional fragmentation does not appreciably affect the broad orbital architecture and mass distribution of the resulting planetary systems \citep[e.g.:][]{deienno19}, we include it in our simulations primarily for the purposes of tracking the evolution of Mercury's core mass fraction (CMF).  The integration timestep in our simulations is 3 days, objects are considered to merge with the Sun at 0.05 au, ejections occur at $Q=$ 100.0 au, and we include additional forces to account for general relativity \citep{saha92}.  All terrestrial objects begin with densities of 3.0 g/cm$^{3}$.  Each individual computation considers four separate regions of the solar system (see figure \ref{fig:cartoon}) derived from previous simulations:

\begin{itemize}
	\item \textbf{(1) Inner ($a<$ 0.7 au) terrestrial disk.} This section of the terrestrial region is alternatively referred to as the ``Mercury-forming region'' throughout our manuscript.  All of our simulations utilize one of two sets of initial conditions  determined to be successful in our previous investigations of Mercury's formation ($\S$ \ref{sect:merc_ics}).  Simulations using the \citetalias{clement21_merc3} model distribute five, 0.05 $M_{\oplus}$ embryos between 0.3-0.7 au with circular, co-planar orbits and randomly assigned semi-major axes that separate them from each other and objects in the outer terrestrial forming disk by $\sim$30-40 $R_{H}$.  In successful realizations, the instability destabilizes this system in a manner such that a single Mercury analog remains as the lone survivor.  Contrarily, runs using the \citetalias{clement21_merc2} initial conditions distribute 20 equal-mass embryos and 200 equal-mass planetesimals between 0.3-0.7 au.  The total mass of planet-forming material is set to 0.5 $M_{\oplus}$ in all models, and equally distributed between the embryo and planetesimal populations.  Thus, inner disk embryos are ten times more massive than the planetesimals.  Eccentricities and inclinations are randomly selected from near-circular Rayleigh distributions, and semi-major axes are assigned in a manner such that the surface density profile of the region increases linearly as $\Sigma \propto r$.
	\item \textbf{(2) Outer (0.7 $<a<$ 4.0 au) terrestrial disk.} The initial conditions for the Venus, Earth and Mars-forming regions used in our new simulations are identical to those employed in \citetalias{clement21_tp} with the notable exception that, here, we remove objects with $a<$ 0.7 au and replace them with the new inner disks described above.  The orbits and masses of our outer disk embryos ($m\geq$ 0.01 $M_{\oplus}$) are the same in all of our simulations, and are taken directly from the GPU-accelerated \citetalias{clement20_psj} simulations of runaway growth at $t=$ 8 Myr (5 Myr after nebular dispersal).  Conversely, each time a simulation restarts we randomly select a new distribution of 861 planetesimals by drawing from the 8 Myr \citetalias{clement20_psj} population of planetesimal ($m<$ 0.01 $M_{\oplus}$) orbits with $a>$ 0.7 au.  The objects are assigned equal-masses such that the cumulative mass of all planetesimals in different radial bins of the outer disk is the same as that of the original GPU models.
	\item \textbf{(3) Giant planet resonant chain.} As in \citetalias{clement18}, \citetalias{clement18_frag} and \citetalias{clement21_tp}, we pre-evolve our giant planet resonant chains in the presence of a planetesimal disk (the primordial Kuiper Belt) up to the point of a close-encounter ($d<$ 3.0 $R_{H}$) between two giant planets before inputing our terrestrial disks ($t_{instb}$ in table \ref{table:ics2}) and evolving the entire solar system through the instability period.  Through this process, we generate a large sample of outer solar systems on the verge of instability (typically the gas giants semi-major axes diverge within 10-100 kyr after restarting these simulations).  The integration timestep for these pre-instability simulations is 50 days.  As in \citetalias{clement21_tp}, we test both a five-planet, 3:2,3:2,3:2 chain (\citetalias{nesvorny12} model) and a five-planet, 2:1,4:3,3:2,3:2 \citetalias{clement21_instb} model where the gas giants' initial eccentricities are $\sim$0.05.  Each time a compute core restarts an instability with a new planetesimal distribution, it also randomly draws a new set of outer solar system initial conditions from our sample. 
	\item \textbf{(4) Primordial Kuiper Belt.}  The planetesimal disk in our initial pre-instability simulations consists of 1,000, equal-mass objects with $a_{nep}+1.5<a<30$ au.  The surface density of the disk falls off as $r^{-1}$, and the total disk mass is set to 35.0 $M_{\oplus}$ in our \citetalias{nesvorny12} instabilities, and 20.0 $M_{\oplus}$ in our \citetalias{clement21_instb} model.  In most cases, the disk is depleted in mass by around a factor of two by the time we input our pre-evolved outer solar system into our terrestrial planet formation simulation.  Thus, our full simulations consider a solar system composed of $\sim$1,500-1,750 objects.
\end{itemize}

\begin{table*}
\centering
\begin{tabular}{c c c c c c c c c c}
\hline
Set & $a_{in}$-$a_{out}$ (au) & $N_{emb}$ & $N_{pln}$ & $M_{emb,tot}$ ($M_{\oplus}$) & $M_{pln,tot}$ ($M_{\oplus}$)  & $t_{instb}$ (Myr) & $N_{sim}$ & Frag\\
\hline
 \multicolumn{8}{c}{\textbf{Comparison models from past work}} \\
\hline
\citetalias{chambers98}/CJS \citepalias{clement18} & 0.5-4.0 & 100 & 1000 & 2.5 & 2.5 & N/A & 50 & N \\
\citetalias{chambers98}/CJS \citepalias{clement18_frag} & 0.5-4.0 & 100 & 1000 & 2.5 & 2.5 & N/A & 100 & Y \\
\citetalias{hansen09}/CJS \citepalias{clement18_frag}  & 0.7-1.0 & 400 & 0 & 2.0 & 0.0 & N/A & 100 & Y \\
\hline
 \multicolumn{8}{c}{\textbf{Comparison instability simulations from past work}} \\
\hline
\citetalias{chambers98}/\citetalias{nesvorny12} \citepalias{clement18} & 0.5-4.0 & 100 & 1000 & 2.5 & 2.5 & 1 & 18 & N \\
\citetalias{chambers98}/\citetalias{nesvorny12} \citepalias{clement18_frag} & 0.5-4.0 & 100 & 1000 & 2.5 & 2.5 & 1 & 22 & Y \\
\citetalias{hansen09}/\citetalias{nesvorny12} \citepalias{clement18_frag} & 0.7-1.0 & 400 & 0 & 2.0 & 0.0 & 10 & 17 & Y \\
\citetalias{clement20_psj}/\citetalias{clement21_instb} \citepalias{clement21_tp} & 0.48-4.0 & 23 & 954 & 2.25 & 2.33 & 5 & 16 & N \\
\hline
 \multicolumn{9}{c}{\textbf{New instability simulations in this paper}} \\
\hline
\citetalias{clement20_psj}+\citetalias{clement21_merc2}/\citetalias{nesvorny12} & 0.3-4.0 & 26 & 1061 & 1.69 & 2.22 & 5 & 16 & Y \\
\citetalias{clement20_psj}+\citetalias{clement21_merc3}/\citetalias{nesvorny12} & 0.3-4.0 & 30 & 861 & 1.82 & 2.09 & 5 & 20 & Y \\
\citetalias{clement20_psj}+\citetalias{clement21_merc2}/\citetalias{clement21_instb} & 0.3-4.0 & 28 & 1061 & 1.69 & 2.22 & 5 & 15 & Y \\
\citetalias{clement20_psj}+\citetalias{clement21_merc3}/\citetalias{clement21_instb} & 0.3-4.0 & 48 & 861 & 1.82 & 2.09 & 5 & 22 & Y \\
\hline	
\end{tabular}
\caption{Summary of initial conditions for complete sets of terrestrial planet formation simulations.  The columns are as follows: (1) the terrestrial (outer$+$inner) disk combination and giant planet instability model used for each simulation set (recall that \citetalias{nesvorny12} refers to the five planet instability model with Jupiter and Saturn in the 3:2 resonance, and \citetalias{clement21_instb} refers to the five planet model with the gas giants in a 2:1 MMR, see the text and figure \ref{fig:cartoon} for definitions of the different disks), (2) the inner and outer edges of the terrestrial forming disk, (3-4) the total number of embryos and planetesimals, (5-6) the total mass of the embryo and planetesimal components, (7) the instability timing in Myr (i.e.: the time giant planets are added to the simulation), (8) the total number of integrations comprising the set, and (9) whether or not (Y/N) the computations were performed using the collisional-fragmentation version of \textit{Mercury} \citep{chambers13}.  The acronyms CJS and EJS refer to Jupiter and Saturn being placed on Circular and Eccentric (modern) orbits, respectively.}
\label{table:ics2}
\end{table*}

\subsection{Constraints}
\label{sect:success}
\begin{table*}
\centering
\begin{tabular}{c c c c c}
\hline
Criterion & Actual Value & Accepted Value & Source & Notes \\
\hline
 \multicolumn{4}{c}{\textbf{Mercury Constraints}} \\
\hline
$M_{Me}/M_{V}$ & 0.0674 & $<$0.2; $>$0.01 & \citetalias{clement21_merc2} crit. \textbf{B} & Calculated as  $\sum M_{i}(a_{i}<a_{V})/M_{V}$\\
$P_{V}/P_{Me}$ & 2.55 & $>$2.25; $q_{V}>Q_{M}$ & \citetalias{clement21_merc2} crit. \textbf{C} & $P_{Me}$ of most massive planet with $a<a_{V}$ \\
$CMF_{Me}$ & 0.7-0.8 & $>$ 0.5 & \citetalias{clement21_merc2} crit. \textbf{D} & See \citetalias{clement18_frag} for CMF calculation \\
\hline
 \multicolumn{4}{c}{\textbf{Other Terrestrial Constraints}} \\
 \hline
$\langle e,i \rangle _{EV}$ & 0.027 & $<$0.055 & \citet{nesvorny21_tp} &  $\langle e,i \rangle _{EV} = (e_{V} + e_{E} + \sin i_{V} + \sin i_{E})/4$ \\
$\Delta a_{EV}$ & 0.28 au & $<$ 0.4 & \citet{nesvorny21_tp} & $\Delta a_{EV} = a_{E} - a_{V}$ \\
$\tau_{\oplus}$ & 30-100 Myr & $>$30 Myr & \citetalias{clement18} crit. \textbf{C} & See \citet{kleine09} \\
$\tau_{Ma}$ & 1-10 Myr & $<$10 Myr & \citetalias{clement18} crit. \textbf{B} & See \citet{Dauphas11} \\
$M_{Ma}/M_{E}$ & 0.107 & $<$0.3; $>$ 0.01 & \citetalias{clement18} crit. \textbf{A} & Calculated as $\sum M_{i}(a_{i}>a_{E})/M_{E}$ \\

$M_{AB,f}/M_{AB,o}$ & $\lesssim$ 0.001 & $<$ 0.02 & \citet{clement18_ab} \\
\hline
 \multicolumn{4}{c}{\textbf{Outer Solar System Constraints}} \\
$e_{55}$ & 0.044 & $>$ 0.022 @ $t_{inst}+$10 Myr & \citetalias{nesvorny12} crit. \textbf{C} & \\
$P_{S}/P_{J}$ & 2.49 & 2.3-2.5 & \citetalias{nesvorny12} crit. \textbf{D} & $<$ 100 kyr between 2.1-2.3  \\
 \hline

\hline
\end{tabular}

\caption{Summary of the success criteria used in our analyses.  The rows are: (1) the ratio of the total planetary mass ($m>$ 0.01 $M_{\oplus}$) remaining inside of Venus' orbit to Venus' mass, (2) the Mercury-Venus orbital period ratio for system's with adequate values of $M_{Me}/M_{V}$, (3) Mercury's final core mass fraction in systems forming analogs, (4) the total eccentricity and inclination excitation of the system's Earth and Venus analogs, (5) the  Earth-Venus semi-major axis spacing, (6-7) the time for Earth and Mars to accrete $90\%$ of their mass, (8) the ratio of the total planetary mass remaining outside of Earth's orbit to Earths' mass, (9) the fraction of asteroidal mass remaining after the instability simulation, (10) the magnitude of Jupiter's fifth eccentric mode and (11) the Jupiter-Saturn period ratio.}
\label{table:crit}
\end{table*}
Our current analyses combine a number of success criteria developed and used in our previous studies of terrestrial planet formation.  We also employ two constraints introduced in \citet{nesvorny21_tp} that eliminate certain redundancies and degeneracies related to commonly-used metrics such as angular momentum deficit and radial mass concentration \citep[e.g.:][]{laskar97,chambers01}.  Table \ref{table:crit} summarizes the 11 important criteria that are relevant to our current study of the early instability.  However, we remind the reader that our outer solar solar system constraints ($e_{55}$ and $P_{S}/P_{J}$) are necessarily satisfied by all simulations as a result of our computational pipeline ($\S$ \ref{sect:sims}).

To analyze a given system, we must first determine which simulation-generated planet (or planets) is an analog of each of the actual terrestrial planets.  We begin by selecting the two most massive terrestrial planets in each system, and denoting them as Earth and Venus analogs.  In contrast to our previous work, we do not analyze simulations that only form a single planet (these infrequent cases are discarded and not included in the discussion of our results).  

Our first three criteria scrutinize Mercury and its dynamical relationship to Venus, and are similar to the metrics employed in \citetalias{clement21_merc2}.  As simulations occasionally form multiple small planets in the Mercury region, rather than picking one and scrutinizing its mass ratio with Venus, we are more interested in the total mass of all planets interior to Venus ($M_{Me}/M_{V}= \sum M_{i}(a_{i}<a_{V})/M_{V}$).  We consider a system to be successful if it finishes with $M_{Me}/M_{V}<$ 0.2.  Systems where Mercury's mass is greater than 3.6 times that of the real planet (0.2 $M_{\oplus}$, i.e.: those with a single Mercury and a massive Venus), as well as those where no Mercury analog originated as an embryo are not included as successful.  While our upper limit on Mercury's mass is rather large, we find it necessary to provide adequate statistics given the small number of Mercury analogs finishing with masses less than twice that of the actual planet in our study.  As a typical Venus analog has a mass of $\sim$0.5-1.0 $M_{\oplus}$ in our models, and each system is initialized with a cumulative planetesimal and embryo mass of 0.25-0.5 $M_{\oplus}$ in the nominal Mercury-forming region, to be successful a system must still lose a significant fraction of the material in the region (typically to merger with the Sun or Venus).  

Another distinctive feature of Mercury and its relationship to Venus is the innermost planets' dynamically isolated orbit \citepalias[explained in detail in][]{clement21_merc3}.  Indeed, even the lowest mass Mercury analogs reported in the terrestrial planet formation literature tend to possess orbits that are far too close to Venus' \citep[e.g.:][]{chambers01,lykawka17,lykawka19}.  To asses our simulations' capacity for replicating this peculiar solar system quality, we require the orbital period ratio between Venus and the largest Mercury analog be greater than 2.25.  While Mercury's moderately excited inclination is also noteworthy for being larger than that of any of the other planets in the solar system, this quality is not particularly challenging to match in numerical simulations \citep[see discussion in:][]{roig16,clement19_merc,nesvorny21_tp}.  Finally, Mercury's distinctive massive iron core \citep[70-80$\%$ of its total mass as estimated by the \textit{MESSENGER} mission:][]{hauck13} presents an interesting constraint for our Mercury analogs.  While it is possible that physical or chemical processes altered the compositions of Mercury's precursor planetesimals \citep[e.g.:][]{ebel11,wurm2013,kruss18,johansen22_merc}, we are keenly interested in understanding the degree to which fragmenting impacts in our simulations might alter Mercury's CMF during the giant impact phase \citep{benz88,asphaug14,ebel17}.  To be considered successful, a Mercury analog (the most massive planet interior to Venus) must finish with a CMF greater than 0.5.  Our methodology for tracking core and mantle transfers in fragmenting collisions is described in detail in \citet{clement19_merc}.  Simply put, we assume that collisional fragments are first produced from the mantle of the projectile.  If the impacting body's mantle is totally eroded, fragments are then produced from the projectile's core, followed by the mantle of the target, and finally its core.  For all simulation sets, we assume that all embryos and planetesimals have CMF $=$ 0.3 around the time of nebular dispersal (we refer to this as ``time zero'' throughout the remainder of the text).  

We use the new constraints introduced in \citet{nesvorny21_tp} as a starting point for constraining our Venus-Earth-Mars systems.  The authors of that paper rightly recognized that the standard metrics employed throughout the literature, namely angular momentum deficit \citep[AMD:][]{laskar97} and radial mass concentration \citep[RMC:][]{chambers01}, possess certain degeneracies that can potentially complicate a rigorous statistical interpretation of simulation outputs.  We analyze these degeneracies in greater detail, and provide further justification for the updated \citet{nesvorny21_tp} metrics in a forthcoming paper  \citep{clement22_tp}.  To briefly summarize these arguments, while AMD and RMC scrutinize the orbital excitation and spacing of the inner planets as a group, the actual solar system is more accurately described by the contrasting properties of the larger and smaller terrestrial worlds.  In particular, Earth and Venus' orbits are remarkably dynamically cold, while those of Mercury and Mars are not.  We scrutinize this property directly by averaging the eccentricities and inclinations of the two most massive planets (Earth and Venus analogs) as:
\begin{equation}
	\langle e,i \rangle _{EV} = \frac{ e_{V} + e_{E} + \sin i_{V} + \sin i_{E} }{4}
	\label{eqn:ei}
\end{equation}
To be successful, a system must attain a value of $\langle e,i \rangle _{EV}$ within a factor of two of the solar system statistic.  Similarly, we measure the Earth-Venus orbital spacing as:
\begin{equation}
	\Delta a_{EV} = a_{E} - a_{V}
	\label{eqn:a}
\end{equation}

Our remaining constraints on Earth and Mars analogs ($\tau_{\oplus}$, $\tau_{Mars}$ and $M_{Ma}/M_{E}$; table \ref{table:crit}) are similar to those employed in our past studies of the early instability scenario.  Here, we define the most massive planet beyond Earth's orbit as the system's Mars analog when calculating $\tau_{Mars}$, and combine all planets in the region when determining $M_{Ma}/M_{E}$.  If no planets form beyond Earth, the system fails both analyses.  More specifically, we require Earth analogs attain $>$90$\%$ of their total mass in $>$30 Myr  \citep[consistent with isotopic dating of the timing of the Moon-forming impact:][]{yin02,wood05,earth,kleine09,rudge10,zube19}, and the largest planet exterior to Earth (Mars analog) accrete 90$\%$ of its mass within the first 10 Myr of the simulation \citep[based off constraints on the planets' accretion derived from analyses of the Martian meteorites:][]{Dauphas11,tang14,kruijer17_mars,costa20_mars}.  However, it is important to note that the precise temporal history of Earth's accretion and the corresponding timing of the Moon-forming giant impact are still debated \citep[e.g.:][]{barboni17,thiemens19}.  As some systems form multiple small planets in the Mars region, in contrast to our previous works, we require the total mass of planets in the Mars-region be no more than 30$\%$ the mass of the system's Earth analog.  As with our Mercury constraints, systems forming no Mars analog that originated as an embryo are not considered to be successful.

It is also important to point out that the large eccentricities of Mars (0.093) and Mercury (0.21) in the solar system are important qualities to be replicated in any formation model.  Past work on terrestrial planet formation and the evolution of the solar system during the giant planet instability have found that these values are common results in simulations \citep{roig16,clement19_merc,lykawka19,desouza21}.  For this reason, we do not include specific constraints on the eccentricities and inclinations of Mercury and Mars in our analyses.  Indeed, the median eccentricity of Mercury (0.17) and Mars (0.12) analogs in our study is in good agreement with those of the real planets.

While our simulation planetesimals are each individually more massive than the estimated cumulative mass of the modern asteroid belt, we are still interested in quantifying the degree of depletion in the belt in our early instability models.  Indeed, if the asteroid belt was heavily populated with planetesimal mass after nebular dispersion as assumed here \citep{hayashi81,bitch15}, it must have subsequently been depleted by some 3-4 orders of magnitude in order to match its present low mass \citep{petit01}.  Though long-term dynamical evolution does transfer mass from the belt to the near-earth asteroid region, it is estimated that the total mass of asteroids has only been reduced by a factor of $\sim$two or so over the past 4 Gyr \citep{minton10}.  While our simulations' number of initial planetesimals in the region (575) hinders us from scrutinizing depletion with the same accuracy employed in other recent high-resolution studies \citep[e.g.:][]{roig15,deienno16,deienno18,clement18_ab}, and the precise initial mass of the belt being poorly constrained \citep[e.g.:][]{ray17sci,dermott18}, as a first-order measure of success we require our systems finish with fewer than 2$\%$ of the initial asteroidal mass.  Thus, assuming an additional depletion factor of $\sim$50$\%$ over the subsequent 4 Gyr, our successful simulations exclusively comprise instabilities capable depleting the belt by at least two orders of magnitude. 

An obvious way in which our classification scheme can break down is in the assessment of a system dominated by three or more massive planets where the most massive two (Earth and Venus analogs) do not inhabit neighboring orbits.  In this case, an overly massive Mars analog might be misclassified as an Earth analog, and the true Earth analog (i.e.: the planet or planets in between the two most massive bodies in the system) would not be considered by any of our analyses.  This could potentially artificially boost our rates of success for $M_{Ma}/M_{E}$ and artificially deflate the number of systems satisfying our $\Delta a_{EV}$ constraint.  However, as the Nice Model disturbance efficiently removes material from the Mars and asteroid belt regions, the two most massive planets in all of our instability models are almost exclusively neighboring planets in the Earth and Venus-region.  Indeed, we verified that only three of our  instability systems finished with a planet with $m>$ 0.3 $M_{\oplus}$ in between the two most massive terrestrial worlds.  Thus, this degeneracy does not appreciably affect the interpretation of the results for our instability models.  Conversely, such configurations occur relatively frequently in our CJS models.  For this reason, we restrict Earth and Venus analogs in CJS models to planets with $a<$ 1.3 au.

It is also possible that a smaller (e.g.: Mars-mass) planet forms between Earth and Venus, and would therefore not be considered in our analyses.  Upon closer inspection we find that, while such outcomes do occur in all of our simulation sets, the additional unclassified planet typically inhabits an unstable orbit that would lead it to eventually merge with Earth or Venus.  As we would not expect such an instability to radically shift $\Delta a_{EV}$ or $\langle e,i \rangle _{EV}$ \citep[see, for example:][]{desouza21}, and we do not impose an upper limit on Earth's accretion timescale ($\tau_{\oplus}$), ignoring such planets should not appreciably affect the interpretation of our results.

\subsection{Comparison Simulation Sets}

Throughout the subsequent sections we compare and contrast our results with those of a number of simulation sets from our past studies (see table \ref{table:ics2}).  As discussed in $\S$ \ref{sect:intro}, outer terrestrial disks (i.e.: beyond 0.7 au) utilized in N-body studies of the various formation scenario can essentially be grouped in two separate categories: extended disks \citepalias{chambers98} and truncated (annulus) disks \citepalias{hansen09}.  While lower order variations on these basic concepts are obviously important for a detailed study of a particular model (e.g.: our updated \citetalias{clement20_psj} extended disk initial conditions for our early instability-model-focused study), it is also important to place the specific results of a study such as ours in the appropriate broader context with clear comparisons to alternative scenarios.   For this reason, we refer to two sets of \citetalias{hansen09} style simulations (one with circular giant planets, and one utilizing a \citetalias{nesvorny12} instability model) initially presented in \citetalias{clement18_frag}.  These models utilized the same fragmentation code \citep{chambers13} described in $\S$ \ref{sect:sims}, and were initialized with 400 equal-mass embryos distributed between 0.7 and 1.0 au.  As annulus models such as these are far more successful at replicating the terrestrial $\langle e,i \rangle _{EV}$ value than extended disk models \citep{nesvorny21_tp}, understanding whether this advantage is maintained when the Nice Model disturbance is accounted for will be an important focus of our subsequent analyses.  Similarly, we also refer to control runs with and without fragmentation (CJS), and the most successful instability simulations from \citetalias{clement18}, \citetalias{clement18_frag} and \citetalias{clement21_tp} in order to understand how the different disk structures and calculation methodologies (namely our collisional fragmentation code) affect our results.  To maximize consistency, we utilize simulations considering 1 and 5 Myr instability delays from those works, and only include simulations that finish with $P_{S}/P_{J}<$ 2.5 and $e_{55}>$ 0.022.  In order to provide a sufficient number of systems for statistical analysis, we combine \citetalias{chambers98}/\citetalias{nesvorny12} style simulations that were initialized with three and four primordial ice giants from both \citetalias{clement18} and \citetalias{clement18_frag} (see figure \ref{fig:cartoon}).


\section{Results}
\begin{table*}
\centering
\begin{tabular}{c c c c c c c c c c}
\hline
Set & $M_{Me}/M_{V}$ & $P_{V}/P_{Me}$ & $CMF_{Me}$ & $\langle e,i \rangle _{EV}$ & $\Delta a_{EV}$ & $\tau_{\oplus}$ & $\tau_{Ma}$ & $M_{Ma}/M_{E}$ & $M_{AB,f}/M_{AB,o}$ \\
\hline
 \multicolumn{9}{c}{\textbf{Comparison control models from past work}} \\
\hline
\citetalias{chambers98}/CJS \citepalias{clement18} & 0 & 12 & N/A & 55* & 8 & 85 & 9 & 0** & 0 \\
\citetalias{chambers98}/CJS \citepalias{clement18_frag} & 6 & 11 & 2 & 83* & 33 & 80 & 33 & 2** & 0 \\
\citetalias{hansen09}/CJS \citepalias{clement18_frag}  & 36 & 5 & 2 & 100* & 56 & 27 & 51 & \textbf{55} & N/A \\
\hline
 \multicolumn{9}{c}{\textbf{Comparison instability simulations from past work}} \\
\hline
\citetalias{chambers98}/\citetalias{nesvorny12} \citepalias{clement18} & 0 & 5 & N/A & 44 & 27 & 93 & 12 & 16 & 33 \\
\citetalias{chambers98}/\citetalias{nesvorny12} \citepalias{clement18_frag} & 13 & 0 & 0 & 50 & \textbf{72} & 90 & 24 & 40 & 50 \\
\citetalias{hansen09}/\citetalias{nesvorny12} \citepalias{clement18_frag} & 17 & 5 & 0 & \textbf{76} & 70 & 20 & 64 & 52 & N/A \\
\citetalias{clement20_psj}/\citetalias{clement21_instb} \citepalias{clement21_tp} &  12 & 6 & N/A & 37 & 37 & 35 & 45 & 18 & 50\\
\hline
 \multicolumn{9}{c}{\textbf{New instability simulations in this paper}} \\
\hline
\citetalias{clement20_psj}+\citetalias{clement21_merc2}/\citetalias{nesvorny12} & \textbf{38} & \textbf{100} & 54 & 46 & 38 & 46 & 45 & 38 & 92 \\
\citetalias{clement20_psj}+\citetalias{clement21_merc3}/\citetalias{nesvorny12}  & 11 & 78 & 35 & 52 & 58 & \textbf{94} & 40 & 35 & 94 \\
\citetalias{clement20_psj}+\citetalias{clement21_merc2}/\citetalias{clement21_instb} & 26 & 80 & 40 & 20 & 13 & 46 & 22 & 46 & \textbf{100} \\
\citetalias{clement20_psj}+\citetalias{clement21_merc3}/\citetalias{clement21_instb} & 13 & 75 & \textbf{75} & 36 & 22 & 77 & \textbf{68} & 22 & 95\\
\hline	
\end{tabular}
\caption{Summary of the percentage of systems in each of our various simulation sets (table \ref{table:ics2}) that satisfy each of our success criteria (table \ref{table:crit}).  Values in bold highlight the most successful simulation set according to each metric.  The columns are as follows: (1) the name of the simulation set, (2) the ratio of the total planetary mass ($m>$ 0.01 $M_{\oplus}$) remaining inside of Venus' orbit to Venus' mass, (3) the Mercury-Venus orbital period ratio for system's with adequate values of $M_{Me}/M_{V}$, (4) Mercury's final core mass fraction in systems forming analogs, (5) the total eccentricity and inclination excitation of the system's Earth and Venus analogs, (6) the  Earth-Venus semi-major axis spacing, (7-8) the time for Earth and Mars to accrete $90\%$ of their mass, (9) the ratio of the total planetary mass remaining outside of Earth's orbit to Earths' mass, and (10) the fraction of asteroidal mass remaining after the instability simulation.  Note that (*) CJS simulations would require a later Nice Model instability to excite the giant planets' eccentricities, and this event would also significantly increase $\langle e,i \rangle _{EV}$ \citep[e.g.:][]{kaibcham16}.  Additionally, we note that (**) the most massive planet formed in \citetalias{chambers98}/CJS disks tends to be near Mars' modern semi-major axis.  Thus, our classification algorithm typically labels it as the Earth analog.  To correct for this, we require Earth and Venus analogs in these simulations have a$<$1.3 au.}
\label{table:results}
\end{table*}

Table \ref{table:results} summarizes the percentages of systems in each of our simulation sets (including both reference models from our previous papers and new calculations performed in this work) that satisfy our various success metrics ($\S$ \ref{sect:success}).  The subsequent sections describe the results of our combined analyses, and are organized as follows: $\S$ \ref{sect:merc} focuses on Mercury, $\S$ \ref{sect:EV} discusses our final Earth-Venus systems, and $\S$ \ref{sect:mars} briefly addresses Mars and the asteroid belt (discussed in more depth in our previous studies: $\S$ \ref{sect:past}).

\begin{figure*}
\centering
\includegraphics[width=.75\textwidth]{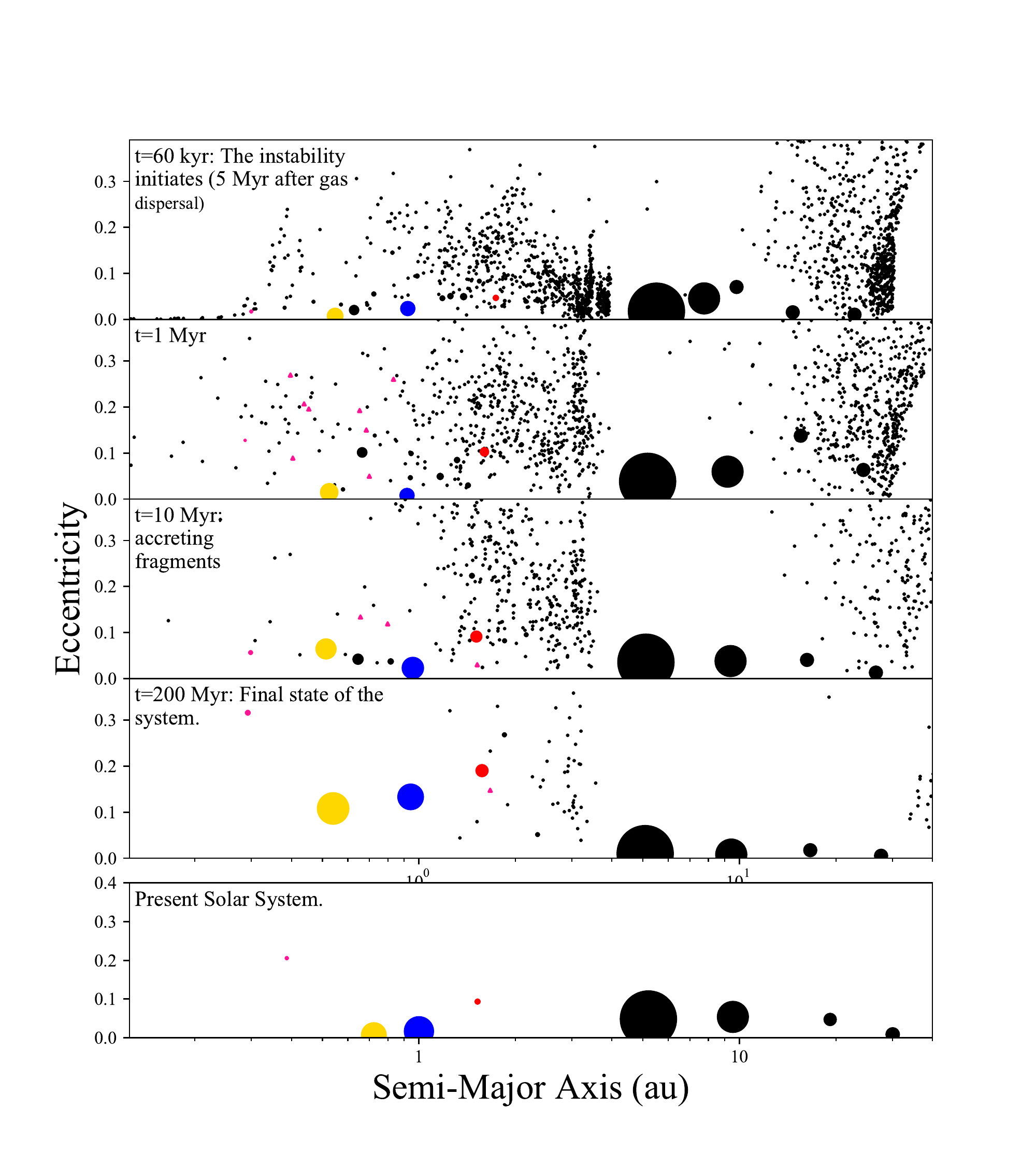}
\caption{Semi-Major Axis/Eccentricity plot depicting the evolution of a successful system in the \citetalias{clement20_psj}+\citetalias{clement21_merc3} batch.  The size of each point corresponds to the mass of the particle (because Jupiter and Saturn are hundreds of times more massive than the terrestrial planets, we use separate mass scales for the inner and outer planets).  For reference, the embryo initial embryos that go on to become Venus, Earth and Mars are color-coded gold, blue and red, respectively.  Collisional fragments are annotated as pink triangles.  The final terrestrial planet masses are $M_{Me}=$ 0.13 ($M_{Me,SS}=$ 0.055) $M_{V}=$ 1.10 ($M_{V,SS}=$ 0.815), $M_{E}=$ 0.85 and $M_{Ma}=$ 0.15 ($M_{Ma,SS}=$ 0.107) $M_{\oplus}$.  The additional important success metrics satisfied by this simulation include: $P_{V}/P_{Me}=$ 2.49, $\Delta a_{EV}=$ 0.40 and $\tau_{\oplus}=$ 50.5 Myr and $M_{Ma}/M_{E}=$ 0.30.  However, Earth and Venus' orbits are overly excited ($\langle e,i \rangle _{EV}=$ 0.21: a common problem in our study, see section \ref{sect:EV_EI}), Mars' formation timescale is slightly too long ($\tau_{Ma}=$ 17 Myr) and the asteroid belt is not sufficiently depleted ($M_{AB,f}/M_{AB,o}=$ 0.040).}
\label{fig:time_lapse}
\end{figure*}

\subsection{Mercury Analogs}
\label{sect:merc}

The primary goal of this study is to assess the compatibility of the new inner disk structures (figure \ref{fig:cartoon}) identified in \citetalias{clement21_merc3} (lone survivor model) and \citetalias{clement21_merc2} (mass-depleted embryo/planetesimal component) with our early instability framework. Unsurprisingly, each simulation set including planet-forming material in the inner, $a<$ 0.7 au region of the terrestrial disk performs remarkably better than any of our previous reference sets when scrutinized against our Mercury-specific constraints: $M_{Me}/M_{V}$, $P_{V}/P_{Me}$ and $CMF_{Me}$.  Indeed, Mercury-Venus systems are essentially ubiquitous in the majority of these models.  In this manner, figure \ref{fig:time_lapse} plots an example temporal evolution of a simulation that utilized our \citetalias{clement20_psj}$+$\citetalias{clement21_merc3}/\citetalias{nesvorny12} initial conditions, and simultaneously satisfied a number of our success criteria.

\subsubsection{Fewer Mercury Analogs from the 2:1 Jupiter-Saturn resonance}

Our simulations that initialize Jupiter and Saturn in a 2:1 MMR with elevated eccentricities (\citetalias{clement21_instb} model) represent a notable exception to the overall trend of Mercury analog generation in our integrations.  Specifically, these new simulation sets possess the highest fraction of system finishing with no mass interior to Venus (10/15 \citetalias{clement20_psj}$+$\citetalias{clement21_merc2} simulations and 14/22 \citetalias{clement20_psj}$+$\citetalias{clement21_merc3} systems).  Contrarily, only 6 total simulations from both of our new instability batches using the \citetalias{nesvorny12} model entirely deplete the Mercury region.  This contrasts with our findings in \citetalias{clement21_tp}.  There, we noted an increased fraction of Mercury analogs formed when Jupiter and Saturn were trapped in the primordial 2:1 resonance.  However, those models did not initialize any planet forming material interior to 0.5 au, and the few Mercury analogs that did form began the simulation as embryos in the Venus-region or, more commonly, the Mars region.  As Jupiter's eccentricity is close to its modern value for the entire duration of these simulations, perturbations from the migration of the $\nu_{5}$ resonance (which begins at low eccentricity and inclination around $a\simeq$ 0.75 au for $P_{J}/P_{S}=$ 2.0) are stronger throughout the duration of the simulation.  While a detailed analysis of this result is beyond the scope of this paper, we remind the reader that reasonable Mercury analogs (according to all three of our metrics) are still obtained in these models (pink-shaded points in the subsequent figures).  However, the majority of these planets have masses that are 2-3 times greater than Mercury's actual mass.  Moreover, in $\S$ \ref{sect:EV} we will argue that the dynamical structure of the Earth-Venus system is also potentially inconsistent with the 2:1 instability model.  Thus, the fact that these models also struggle to produce Mercury analogs further strengthens our arguments in favor of the 3:2 (\citetalias{nesvorny12}) model.

\subsubsection{Matching the Mercury-Venus period ratio with \citetalias{clement21_merc3} disks}

\begin{figure*}
	\centering
	\includegraphics[width=.85\textwidth]{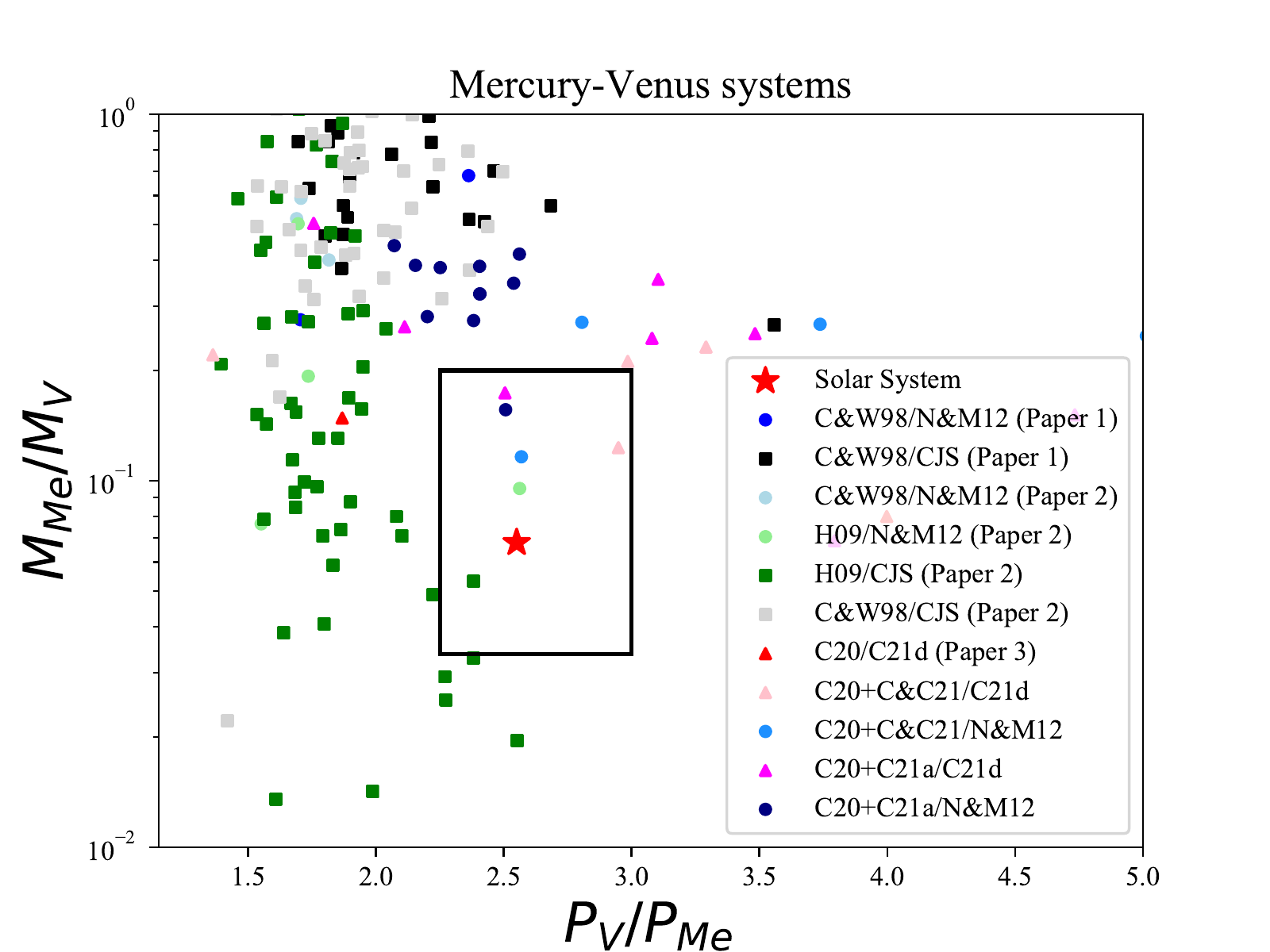}
	\caption{Final values of $M_{Me}/M_{V}$ versus the Mercury-Venus period ratio, $P_{V}/P_{Me}$ for all Mercury-analogs formed in our present study (see explanation in $\S$ \ref{sect:success}) compared with those from our past studies.  Simulations that do not include a giant planet instability (CJS) are annotated with squares.  Instabilities where Jupiter and Saturn begin in a 3:2 MMR (\citetalias{nesvorny12} model) are plotted in shades of blue, while shades of red or pink triangles indicate 2:1 (\citetalias{clement21_instb}) models.  All annulus (\citetalias{hansen09}) simulations are plotted in shades of green.  The red star indicates the solar system value.  The black square highlights the region bounded between a period ratio of 2.25 (our success criterion) and 3.0 (not to exceed the 3:1 MMR), and a mass ratio of 0.034 (half the current value to ensure stranded planetesimals are not included) and 0.2 (our success criterion).}
	\label{fig:merc_ven}
\end{figure*}

Figure \ref{fig:merc_ven} plots the distribution of Mercury-Venus mass and period ratios attained in all of our new simulations, compared with selected reference models from our past work (table \ref{table:ics2}) out to a period ratio of 5.0.  To highlight the most successful analog systems, we plot a box that bounds the systems finishing with 2.25 $< P_{V}/P_{Me} <$ 3.0 and 0.034 $< M_{Me}/M_{V} <$ 0.2.  These limits are based on our success criteria for the values (table \ref{table:crit}), the Mercury-Venus 3:1 MMR, and half the current value of $M_{Me}/M_{V}$. Given the number of outer solar system constraints already applied to these simulations prior to analysis, it is encouraging that there is at least one reasonable Mercury analog in this box for each instability model (3:2 and 2:1) and inner disk structure (\citetalias{clement21_merc3}, \citetalias{clement21_merc2} and \citetalias{hansen09} annulus). 

While some of our Mercury-Venus systems exceed $P_{V}/P_{Me}=$ 5.0 (discussed below), we exclude these from this figure as they obviously represent poor outcomes.  Though it is clear from the distribution of plotted values that the outcomes of our new instability simulations (blue and pink shaded points) span the complete range of possible $P_{V}/P_{Me}$ values, they also tend to finish with $M_{Me}/M_{V} \simeq$ 0.1-0.3; slightly greater than the solar system value of 0.067.  While this is clearly a marked improvement from all of the extended disk \citepalias{chambers98} models from \citetalias{clement18} and \citetalias{clement18_frag}, it is interesting that the \citetalias{hansen09} (annulus) simulations more consistently yield Mercury analogs with the appropriate mass.  However, with the exception of one instability model (light green point on top of the red star in figure \ref{fig:merc_ven}; our best Mercury-Venus system), no annulus model finishes with $P_{V}/P_{Me}$ in excess of the solar system value.

\begin{figure*}
	\centering
	\includegraphics[width=.75\textwidth]{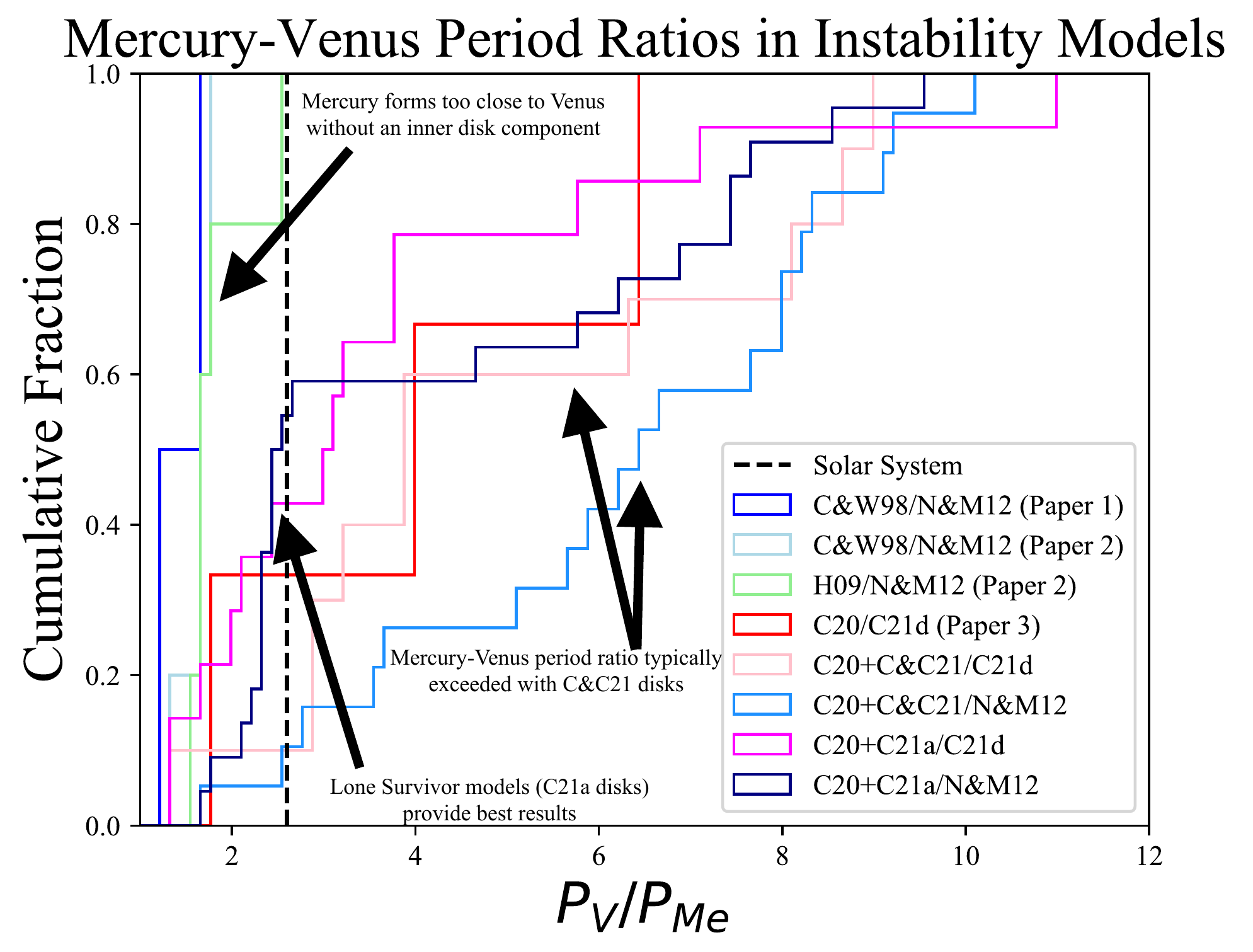}
	\caption{Cumulative distribution of final $P_{V}/P_{Me}$ values for Mercury-Venus systems formed in giant planet instability models.  The black dashed vertical line indicates the solar system value.  Instabilities where Jupiter and Saturn begin in a 3:2 MMR (\citetalias{nesvorny12} model) are plotted in shades of blue, while shades of red or pink indicate 2:1 (\citetalias{clement21_instb}) models.  All annulus (\citetalias{hansen09}) simulations are plotted in shades of green.}
	\label{fig:prat_inst}
\end{figure*}

To further expound on the challenges involved in replicating the Mercury-Venus period ratio in our models, we plot the cumulative distribution of all final system $P_{V}/P_{Me}$ statistics for all of our instability simulation batches in figure \ref{fig:prat_inst}.  To bolster statistics in this plot, we include simulations that finish with 2.5 $<P_{J}/P_{S}<$ 2.8 that are not considered in any of our other analyses.  It is again clear from this figure that our reference models from \citetalias{clement18} and \citetalias{clement18_frag} that truncate the inner terrestrial disk around the vicinity of Venus' modern orbit struggle to produce Mercury analogs in general.  Moreover, the rare analogs themselves are systematically too close to Venus.  Furthermore, a second trend emerges upon closer inspection of the differences between our two inner disk models (\citetalias{clement21_merc3} and \citetalias{clement21_merc2}).  While the majority of our \citetalias{clement21_merc2} simulations that place a collection of 20 embryos and 200 planetesimals in the inner disk finish with Mercury-Venus period ratios significantly in excess of the actual value, the distribution of results for our \citetalias{clement21_merc3} systems cluster more tightly around $P_{V}/P_{Me}=$ 2.6 (solar system value) in figure \ref{fig:prat_inst}.  Despite our \citetalias{clement21_merc2} simulation sets possessing slightly improved success rates for the first three metrics listed in table \ref{table:results}, for this reason, we assess our \citetalias{clement20_psj}$+$\citetalias{clement21_merc3} initial conditions to be the most successful in terms of their ability to generate the authentic Mercury-Venus system.

In our initial study of the ``lone survivor'' scenario where $\sim$4-6 Mars-mass embryos interior to Venus destabilize and leave behind a single Mercury analog though a sequence of fragmenting collisions \citepalias{clement21_merc3} we exclusively modeled the giant planets on their current orbits.  A major shortcoming of those simulations was a tendency of the final Mercury-Venus analogs to have excessively large orbital period ratios, however we speculated that this might be resolved if the event transpired during giant planet migration.  Our new simulations confirm this suspicion.  Specifically, 13/20 total simulations in our \citetalias{clement20_psj}$+$\citetalias{clement21_merc3}/\citetalias{nesvorny12} set (lone survivor model with 3:2 version of the Nice Model) form Mercury analogs with 2.0 $< P_{V}/P_{Me} <$ 3.0.  While only one of these analogs simultaneously satisfies our relatively strict constraint on Mercury's mass, it is clear from figure \ref{fig:merc_ven} that very few systems are successful in this manner generally (we note that there is still a large amount of uncertainty in collisional fragmentation models that might also account for some of Mercury's mass depletion, see $\S$ \ref{sect:cmf} for additional discussion).  Moreover, an additional two of these 13 systems finish with two Mercury analogs, neither of which is more massive than 0.2$M_{V}$.  Thus, a late collision with Venus would be the only event separating these simulations from success.

To better understand how our new instability simulations provide better matches to $P_{V}/P_{Me}$ than our original \citetalias{clement21_merc3} models, we inspect the frequency at which each specific proto-planet survives to become the system's Mercury analog, as well as the average time of, and reason for the loss of the other four embryos.  In our former simulations that only considered the evolution of 5 proto-planets interior to the seven other planets on their modern orbits, the innermost embryo survived as the system's final Mercury analog 44$\%$ of the time.  Contrarily, this was only the case in 32$\%$ of our  \citetalias{clement20_psj}$+$\citetalias{clement21_merc3}/\citetalias{nesvorny12} instability simulations.  In fact, we also observe instances where a rouge planetesimal or embryo from the Venus-forming region is scattered inward onto a Mercury-like orbit.  Additionally, without perturbations from the instability included, it is far more likely for all proto-planets to combine into a single, overly massive Mercury analog.  In \citetalias{clement21_merc3} we observed no cases where the innermost embryo merged with Venus or Earth, and only a single instance in over 200 simulations where the second proto-planet merged with one of the fully-formed terrestrial planets.  Contrarily, 9 of the embryos closest to the Sun in our 20 instability simulations combine with Venus; thus reducing the probability of forming a Mercury analog with an excessively large value of $P_{V}/P_{Me}$.  Thus, we argue that the incorporation of an instability model improves the likelihood that Mercury forms at the correct semi-major axis by removing excessive embryos with low semi-major axes.  

\subsubsection{Enhanced $CMF_{Me}$ in instability simulations}
\label{sect:cmf}

\begin{figure}
	\centering
	\includegraphics[width=.5\textwidth]{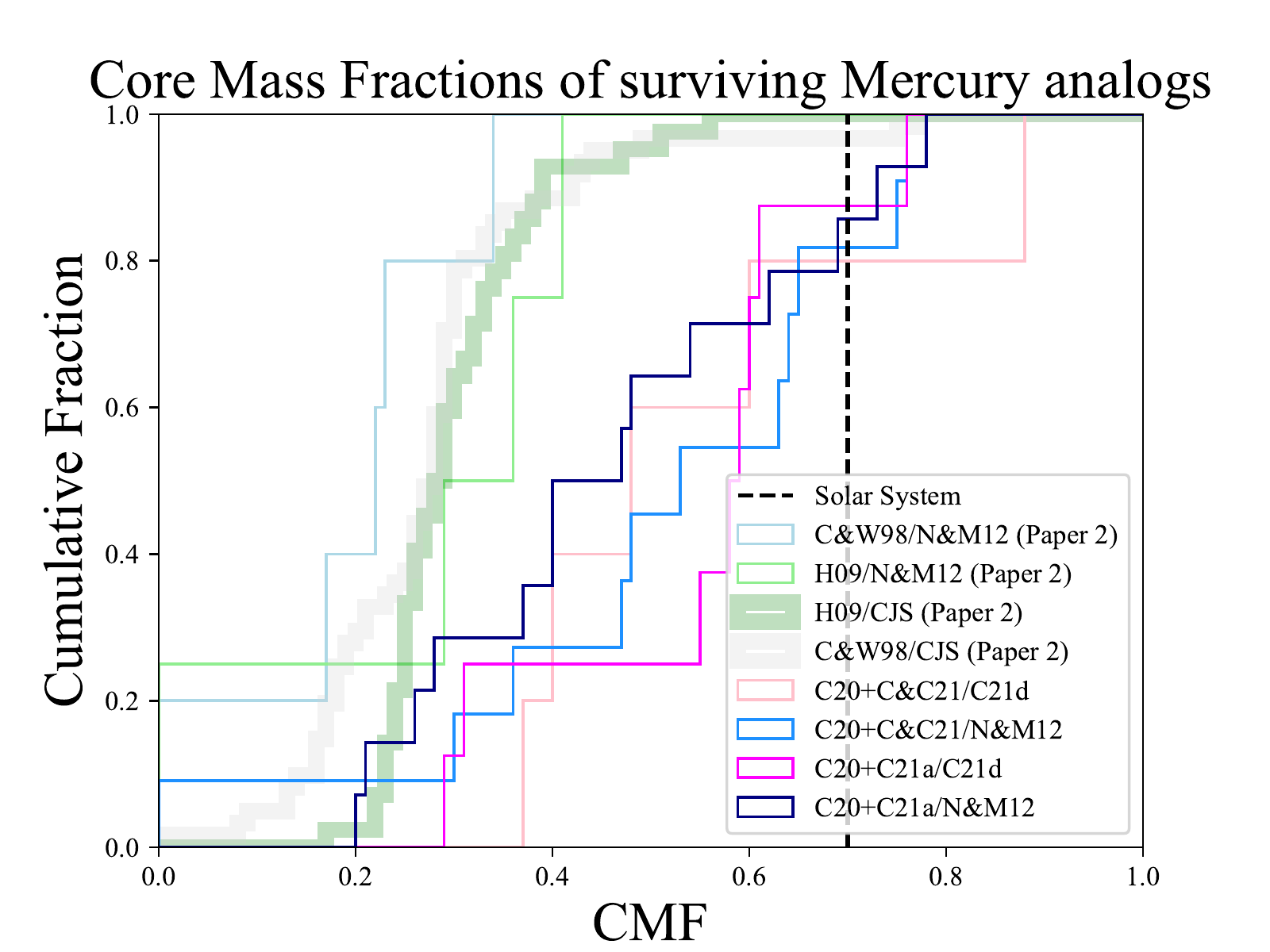}
	\caption{Cumulative distribution of final CMFs of Mercury analogs formed in simulation sets utilizing collisional fragmentation \citep{chambers13}.  The black dashed vertical line indicates Mercury's actual approximate inferred CMF \citep{hauck13}.  Instabilities where Jupiter and Saturn begin in a 3:2 MMR (\citetalias{nesvorny12} model) are plotted in shades of blue, while shades of red or pink indicate 2:1 (\citetalias{clement21_instb}) models.  All annulus (\citetalias{hansen09}) simulations are plotted in shades of green.}
	\label{fig:cmf}
\end{figure}

Figure \ref{fig:cmf} plots the cumulative distribution of Mercury CMFs for all of our fragmentation simulations (reference models from \citetalias{clement18_frag} and new runs from this work).  Notably, regardless of the specific structure of the terrestrial disk, instability simulations produce a greater fraction of high-CMF Mercury alongs than the corresponding CJS simulations.  This is unsurprising given that the instability-induced eccentricity excitation of the terrestrial disk provides high-speed collisions that have the potential to fall in the fragmentation regime.  Coupled with the aforementioned improved $P_{V}/P_{Me}$ results for certain simulation sets, this result illustrates how an early instability might provide a compelling explanation for Mercury's peculiar orbit and internal structure.  

It is also clear from figure \ref{fig:cmf} that the inferred value of $CMF_{Me}$ ($\sim$0.7) lies outside the range of values produced in any of our \citetalias{hansen09} annulus models (both with and without giant planet migration).  This is largely a consequence of the fact that Mercury essentially forms as a stranded embryo in these models.  Once Mercury is ejected from the annulus region, its accretion is essentially over.  Thus, these Mercury analogs tend to have less overall time and correspondingly fewer accretion events to alter their CMFs.  While it is certainly possible that the Mercury's precursor planetesimals and embryos were altered in CMF via chemical or physical processes prior to its ejection from the region, it would be difficult to explain how this process did not also affect the Earth or its precursors since both planets must necessarily originate in the same annulus.  However, as the nature and size of Venus' core remains largely unconstrained \citep[e.g.:][]{venus_core1,venus_core2}, future exploration (e.g.: DAVINCI$+$ and VERITAS) will undoubtedly be key in providing improved constraints for our terrestrial formation models.

In general, similar fractions of Mercury analogs in our new instability simulations (rightmost blue and pink lines in figure \ref{fig:cmf}) attain CMFs in excess of 0.5.  It is important to note that the percentages provided in the fourth column of table \ref{table:results} report the fraction of all simulations in the respective set that finish with a high-CMF Mercury analog.  Thus, sets of initial conditions that struggle to form such planets in general (specifically our 2:1 \citetalias{clement21_instb} instabilities) finish with lower scores, even if the distribution of CMFs in figure \ref{fig:cmf} is similar to those of the more successful batches.  

\begin{figure}
	\centering
	\includegraphics[width=.5\textwidth]{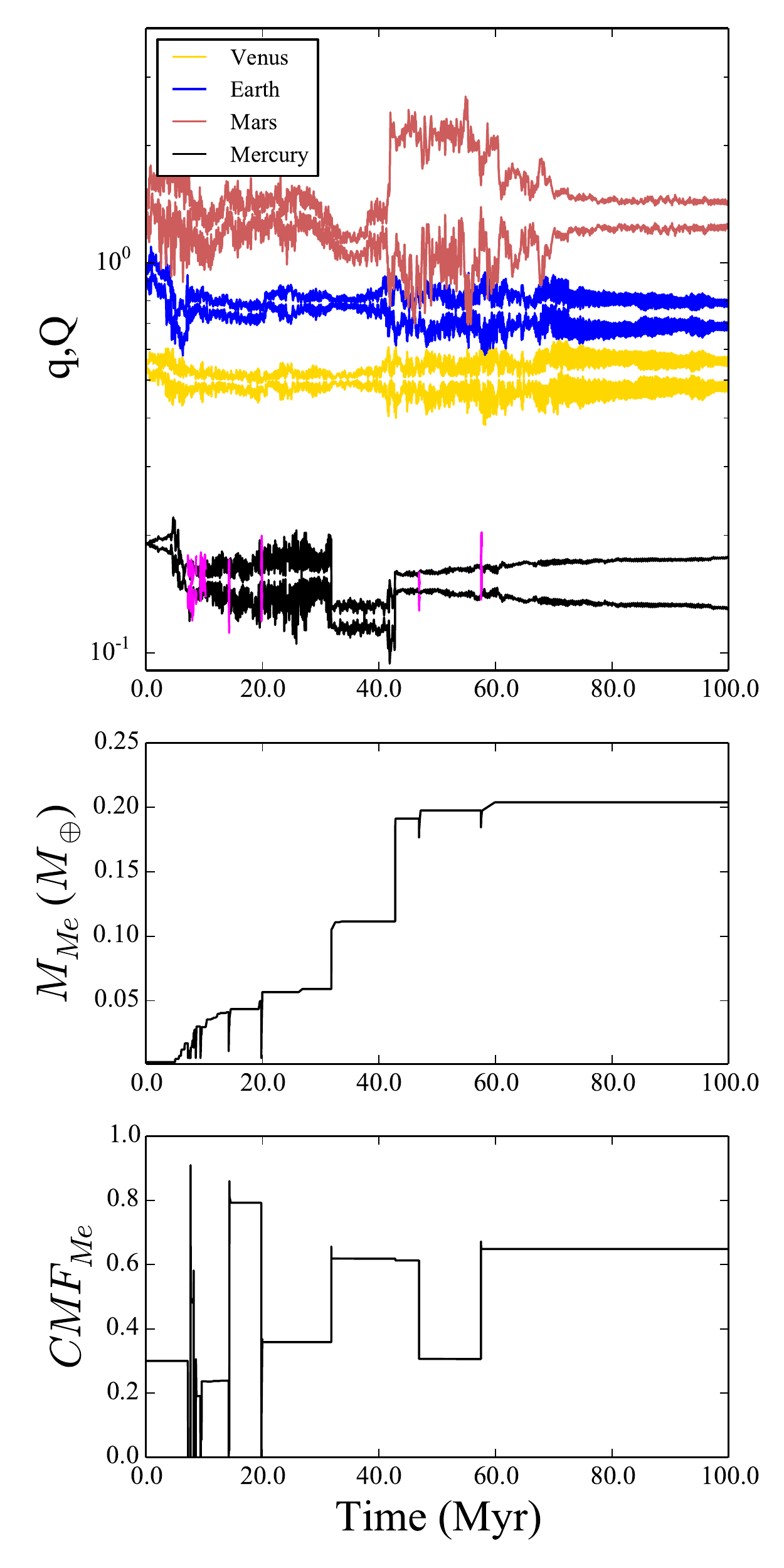}
	\caption{Top panel: Perihelia and aphelia of the four terrestrial planets from a simulation in our \citetalias{clement20_psj}$+$\citetalias{clement21_merc2}/\citetalias{nesvorny12} set.  The pink lines plot the q/Q of collisional fragments ejected from the Mercury analog (black line) that alter its CMF during its growth.  Middle and Bottom panels: Mass and CMF evolution of the same Mercury analog, respectively.}
	\label{fig:qaq}
\end{figure}

While the CMFs of Mercury analogs in our \citetalias{clement21_merc3} simulations that only consider five proto-planets in the inner disk are typically altered in a series of fragmenting interactions between a pair of specific embryos, the CMF evolution of inner disk objects in our \citetalias{clement21_merc2} disks is often exceedingly complex.  Figure \ref{fig:qaq} plots one example evolution for a system from our \citetalias{clement20_psj}$+$\citetalias{clement21_merc2}/\citetalias{nesvorny12} batch that satisfies all but one ($\tau_{Ma}$) of our success criteria.  In total, 39 collisional fragments (color-coded pink in the top panel) are ejected from this analog over the duration of its growth.  One intriguing aspect of this analog's evolution is the fact that, prior to being permanently enhanced in CMF, the proto-Mercury is involved as the smaller object in five high-velocity, erosive glancing collisions and re-ejected as a mantle-only fragment.  As the original projectile in such interactions is considered the ``first fragment'' by our algorithm, it is always assigned the initial mantle portion of the ejected material.  Similarly, several near equal-mass accretion events involving other fragments composed of mostly core-material radically alter the analog's CMF in the positive direction.

It is important to consider how the random assignment of core and mantle material to the particles produced in fragmenting collisions when post-processing our results can artificially enhance or reduce the final CMF of a specific Mercury analog in a specific simulation.  If a hypothetical final surviving, high-CMF planet in the inner disk originated as a core-only fragment after a catastrophically destructive impact, it is easy to see how the system might have been unsuccessful if we had simply assigned mantle material to that specific fragment, rather than core-material.  Thus, we are far more interested in the distribution of final CMFs in figure \ref{fig:cmf} than we are in, for example, the degree of mutual exclusivity between high $CMF_{Me}$ and other constraints in individual systems.

\subsection{Earth-Venus System}
\label{sect:EV}

\subsubsection{Earth's growth and the Moon-forming impact}
As demonstrated in table \ref{table:results} and our past studies of the early instability ($\S$ \ref{sect:past}), prolonging Earth's accretion ($\tau_{\oplus}>$ 30 Myr) is not a challenging constraint to satisfy.  While Earth grows rapidly if the planet-forming material is concentrated in an annulus \citep[such models often argue that a five planet inner solar system was rapidly assembled, and then survived in a quasi-stable state until the Moon-forming impact, e.g.:][]{johansen21,broz21,izidoro22_ss}, accretion in our extended disk models is typically more prolonged.

\begin{figure}
	\centering
	\includegraphics[width=.5\textwidth]{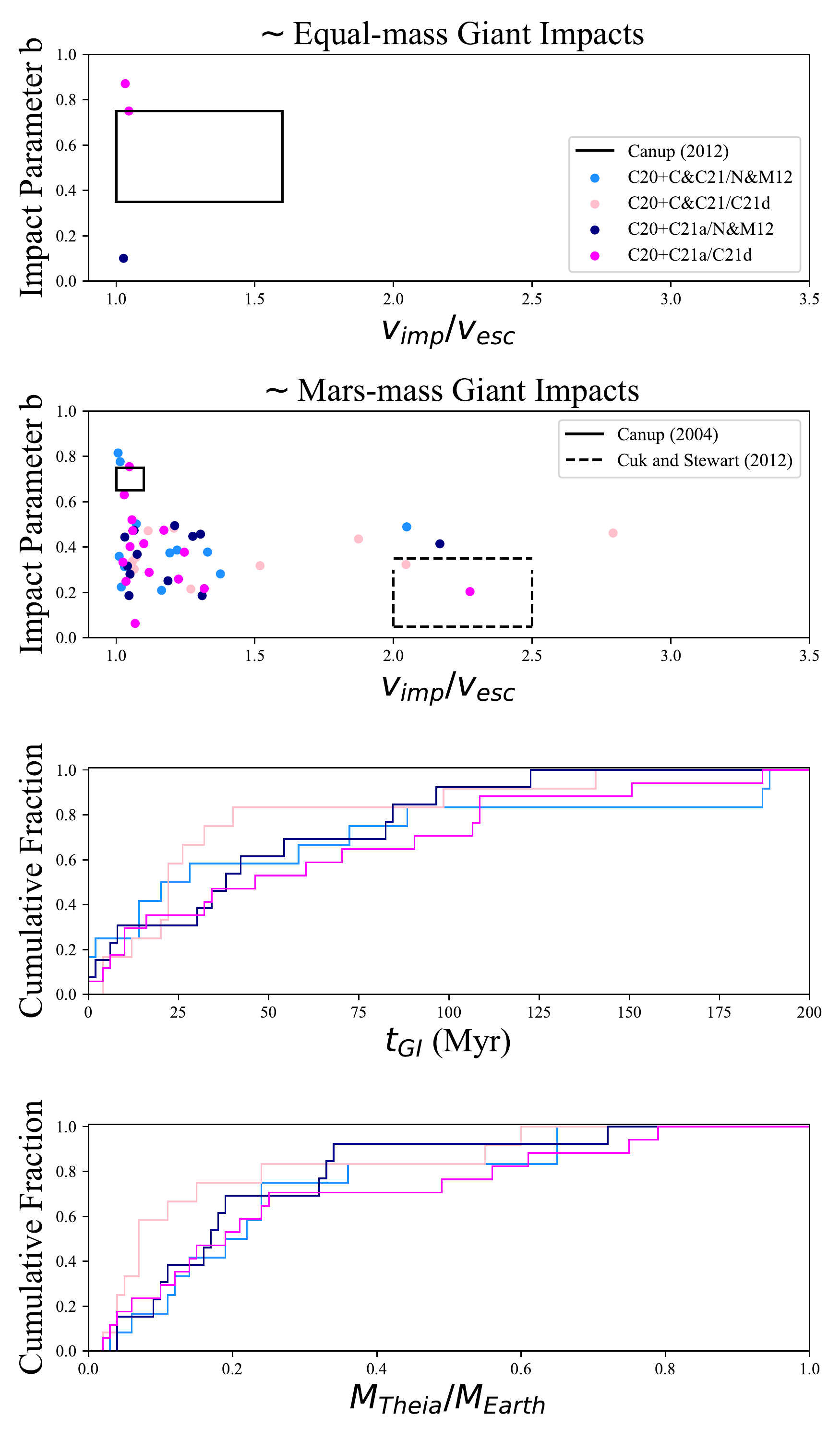}
	\caption{The top panel plots impact parameters ($b$) and velocities ($v_{imp}/v_{esc}$) for all roughly equal-mass ($M_{Theia}/M_{Earth}>$ 0.7) Moon-forming impacts (defined here as the final collision between Earth and an embryo) in our new instability simulations.  The solid square denotes the preferred parameters from the analysis of \citet{canup12}.  The second panel is similar to the top panel, except here we only display final giant impacts where $M_{Theia}/M_{Earth}<$ 0.5.  The solid lines mark the preferred geometry of \citet{canup04}, and the dashed lines indicate the successful impacts from the rapid-rotating model of \citet{cuk12} (see text).  The third and fourth panels plot the cumulative distributions of the final giant impact times ($t_{GI}$) and Theia:Earth mass ratios in our simulation sets.  Instabilities where Jupiter and Saturn begin in a 3:2 MMR (\citetalias{nesvorny12} model) are plotted in shades of blue, while shades of red or pink indicate 2:1 (\citetalias{clement21_instb}) models.}
	\label{fig:moon}
\end{figure}

Figure \ref{fig:moon} analyzes the nature of the final giant impacts on Earth analogs (the nominal Moon-forming event) in our various instability simulation sets.  While previous studies have scrutinized the connection between the types of impacts produced N-body studies of terrestrial planet formation \citep[e.g.:][]{kaibcowan15,quarles15,desouza21,woo22} and the particular types of Moon-forming events found to be consistent with orbital and chemical constraints via hydrodynamical modeling \citep[e.g.:][]{canup04,cuk12,canup12,reufer12,lock17,carter20} we did not analyze our simulations in this manner in our past studies that were more focused on Mars' formation.

In general, there are two types of proposed impact scenarios that might have led to the formation of the Earth-Moon system: those where the impactor, Theia, has a mass approximately equal to that of Mars \citep[the so-called canonical scenario, e.g.:][]{benz86,canup04,asphaug14_moon} and those preferring a collision between roughly equal-mass objects \citep[e.g.:][]{canup12}.  As our \citetalias{clement20_psj} initial outer terrestrial disks are dominated by three, $m\simeq$ 0.3 $M_{\oplus}$ embryos at time zero, we naively expected equal-mass Moon-forming events to be common occurrences in our simulations.  However, it is clear from the upper two panels that this is not the case.  Indeed, equal-mass large accretion events (defined here as $M_{Theia}/M_{Earth}>$ 0.7) are far more common on Venus early in our simulations, while Earth's accretion is typically more prolonged, and involves the addition of multiple Mercury-Mars mass objects \citepalias[see additional discussion in][]{clement21_tp}.  Indeed, over $\sim$80$\%$ of the final giant impacts in all of our different instability simulations have projectile:target mass ratios ($M_{Theia}/M_{Earth}$ in the bottom panel of figure \ref{fig:moon}) less than $\sim$0.2.

In the canonical model for the formation of the Moon \citep[e.g.:][]{hartmann75,cameron76,benz86}, the angular momentum of the Earth-Moon system is considered to be a conserved quantity.  High resolution simulations of this scenario utilizing smoothed particle hydrodynamics (SPH) codes indicate that a low-velocity ($v_{imp}/v_{esc}\lesssim$ 1.05) collision at an oblique angle ($\sim$45\degr) is most consistent with constraints from the modern systems' mass partitioning and total angular momentum \citep{canup01,canup04b,canup04}.  A slight variation on this model was proposed in \citet{reufer12}, where the authors advocate for a slightly more energetic ($v_{imp}/v_{esc}\simeq$ 1.2) hit-and-run impact that leaves behind a Moon analog predominantly derived from proto-Earth material.  While such a scenario potentially requires Theia be more water-rich and possibly originate in the main belt \citep{jackson18}, given the similarities between the \citet{reufer12} and \citet{canup04} models we mostly consider them jointly when analyzing our simulations.  It is clear from the distributions plotted in the second panel of figure \ref{fig:moon} that similar events are common in our models.  However, it is important to interpret the rather low rate at which each model produces the precise impact geometry (i.e.: yielding a point inside the box) within the appropriate context of the highly stochastic giant impact phase.  Moreover, while the the distribution of Moon formation times ($t_{GI}$) in the third panel of figure \ref{fig:moon} includes many early events ($t\lesssim$ 30 Myr) that are potentially inconsistent with isotopically inferred ages \citep[e.g.:][]{kleine09}, impacts that are good analogs to the preferred canonical impact disproportionately occur later in simulations.  Indeed, the median value of $t_{GI}$ for 0.5 $<b<$ 0.85, $M_{Theia}/M_{Earth}<$ 0.5 and $v_{imp}/v_{esc}<$ 1.25 is 63.1 Myr.

A variation of the canonical, Mars-mass impactor model was proposed by \citet{cuk12}.  The authors proposed a high-speed impact involving a rapidly rotating proto-Earth and an impactor that strikes at an orientation retrograde to Earth's spin.  Furthermore, the model does not require the system angular momentum be conserved, and instead argues that the rapidly rotating proto-Earth-Moon system can be spun-down through the evection resonance with the Sun.  The dashed lines in the second panel of figure \ref{fig:moon} denote the preferred impact parameters from the \citet{cuk12} analysis, and demonstrate feasibility of such a Moon-forming event occurring within the early instability framework.  It is interesting that similarly energetic interactions occur with the greatest frequency in our \citetalias{clement21_instb} instability models that initialize Jupiter and Saturn in a 2:1 MMR with elevated eccentricities.  This is most likely the consequence of an increased tendency of terrestrial over-excitation in these instability scenarios.  Indeed, both \citetalias{clement21_instb} sets have significantly lower $\langle e,i \rangle _{EV}$  success rates (20 and 36$\%$; table \ref{table:results}) than their \citetalias{nesvorny12} instability model counterparts (46 and 52$\%$).  We elaborate further on these trends in the subsequent section.

 In summary, while our simulations produce reasonable matches to the proposed impact geometries of a number of Moon-formation scenarios in the literature, the most common final giant impact on Earth analogs in our models most closely matches the canonical impact scenario of \citet{canup04}.  While perturbations from the Nice Model instability do excite orbits in the terrestrial region, exotic configurations such as the equal-mass model of \citet{canup12} and the fast-spinning scenario of \citet{cuk12} are not regular outcomes in our simulations.  However, these results should be taken in the appropriate context given the fact that our analyses are limited by small number statistics.

\subsubsection{Replicating Earth and Venus' cold orbits}
\label{sect:EV_EI}

\begin{figure}
	\centering
	\includegraphics[width=.5\textwidth]{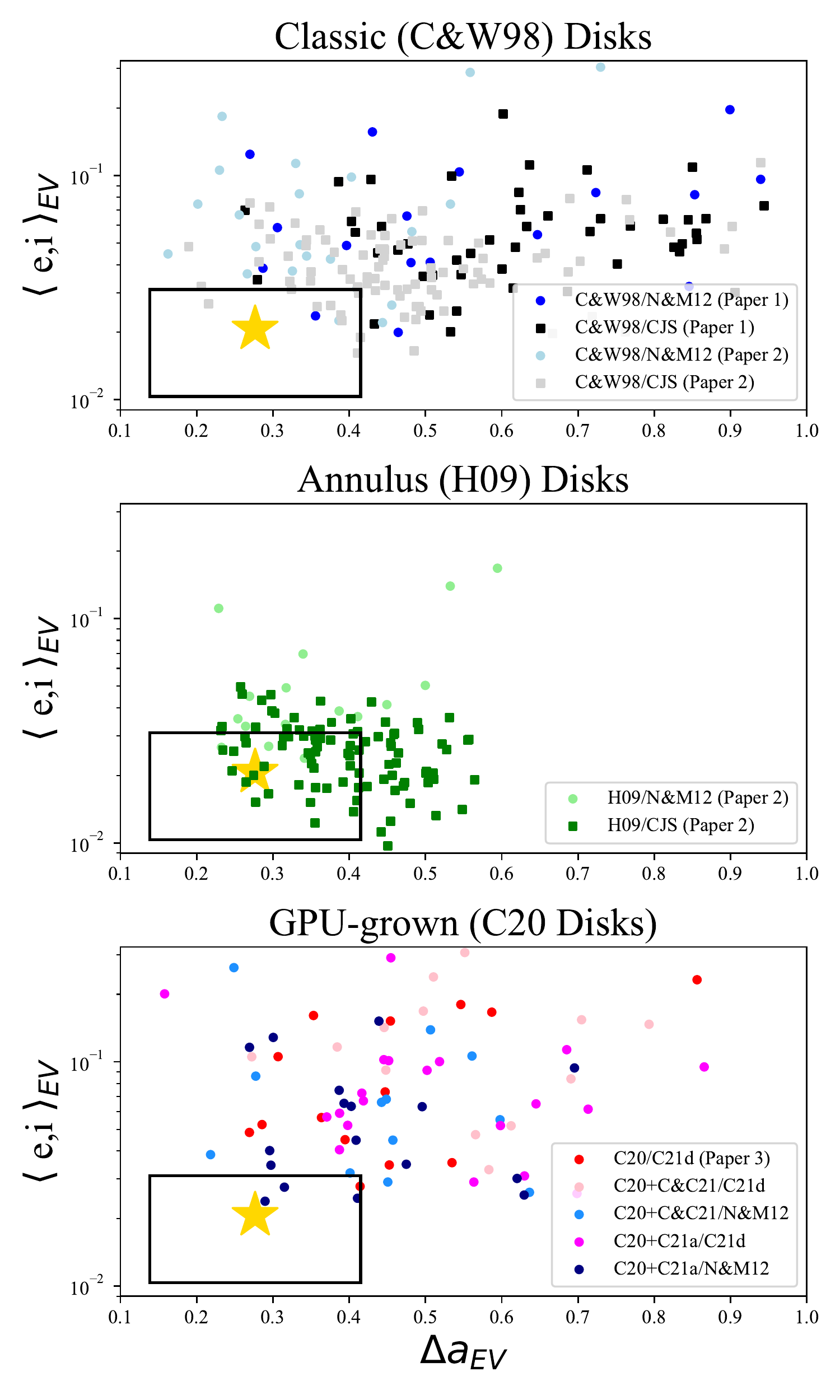}
	\caption{Distribution of Earth-Venus system dynamical success criteria ($\langle e,i \rangle _{EV}$ versus $\Delta a_{EV}$) for our various simulation sets.  The top two panels compile simulations from our past studies of the early instability.  Specifically, the top panel combines instability and static giant planet simulations performed using classic disk initial conditions \citepalias{chambers98}, and the middle panel displays similar data for annulus \citepalias{hansen09} disk conditions.  The bottom panel combines our most successful simulations from \citetalias{clement21_tp} and those from our present study.  As in figure \ref{fig:merc_ven}, simulations that do not include a giant planet instability (CJS) are annotated with squares.  Instabilities where Jupiter and Saturn begin in a 3:2 MMR (\citetalias{nesvorny12} model) are plotted in shades of blue, while shades of red or pink indicate 2:1 (\citetalias{clement21_instb}) models.  All annulus (\citetalias{hansen09}) simulations are plotted in shades of green.  The yellow star indicates the solar system value, and the solid lines enclose systems that finish with both metrics between 0.5 and 1.5 times the solar system values (note that these limits to not correspond to our success criteria utilized in tables \ref{table:crit} and \ref{table:results}, and are instead selected to highlight the scarcity of solar system-like results).}
	\label{fig:ei_ev}
\end{figure}

Figure \ref{fig:ei_ev} compares the distributions of final $\langle e,i \rangle _{EV}$ and $\Delta a_{EV}$ values for Earth-Venus systems formed in all simulations utilizing our three primary disk configurations: \citetalias{chambers98}, \citetalias{hansen09} and \citetalias{clement20_psj}.  It is obvious from the plotted distributions that the solar system values (yellow stars) lie at the extreme of the range of simulation-generated values.  While the annulus models plotted in the middle panel produce the greatest number of systems with terrestrial-like low-$\langle e,i \rangle _{EV}$ values and small Earth-Venus orbital spacings, the majority of these successful systems are produced in simulations that do not include a giant planet instability model.  As discussed in the introduction, \textit{all terrestrial planet formation models must account for the effects of the giant planet instability.}  While the primary difference between the instability and CJS models in all panels of figure \ref{fig:ei_ev} is hotter $\langle e,i \rangle _{EV}$ distributions ($\Delta a_{EV}$ is not particularly affected by the instability), reasonable analogs are still produced in simulations that include giant planet migration.  

Only one of our reference instability simulations from \citetalias{clement18} and \citetalias{clement18_frag} that utilized the \citetalias{chambers98} disk finishes in the black box around the yellow star in the upper panel of figure \ref{fig:ei_ev} that denotes outcomes falling between 0.5 and 1.5 times the solar system values.  While we cannot rule out the \citetalias{chambers98} disk simulations from \citetalias{clement18_frag} as incompatible with the early instability scenario on these grounds, it is clear that our instability simulations utilizing the \citetalias{hansen09} and our new \citetalias{clement20_psj} disks provide more compelling results.  Clearly, the distribution of outcomes for the \citetalias{hansen09} instability simulations cluster more tightly around the solar system value.  However, the fact that the \citetalias{clement20_psj} disks produce reasonable results as well makes it difficult to argue that one disk structure represents the authentic state of the inner solar system around the time of nebular gas dispersal, while the other does not.  Indeed, our best \citetalias{clement20_psj} instability sets in this paper satisfy our constraints for $\langle e,i \rangle _{EV}$ and $\Delta a_{EV}$ around 50$\%$ of the time. as compared to $\sim$70$\%$ for the annulus instability runs.  Although our analyses in $\S$ \ref{sect:merc} disfavor the \citetalias{hansen09} disks because they produce poor $P_{V}/P_{Me}$ values, this might be improved if a concentrated annulus outer terrestrial disk structure were combined with one of our inner disk configurations (\citetalias{clement21_merc3} or \citetalias{clement21_merc2}).  

Nevertheless, for our current purposes it seems reasonable to simply conclude that the \citetalias{hansen09} and \citetalias{clement20_psj} disks are both compatible with orbital constraints from the Earth-Venus system when they are evolved through the Nice Model instability.  However, our work reinforces the fact that Earth and Venus' cold orbits are extremely challenging to replicate with embryo accretion models, and we note two important caveats on our overall findings:
\begin{itemize}
	\item \textbf{\textit{Collisional fragmentation likely played a minor, albeit important role in Earth and Venus' dynamical evolution:}} It is also clear from figure \ref{fig:ei_ev} that our best results in terms of simultaneously replicating both $\langle e,i \rangle _{EV}$ and $\Delta a_{EV}$ occur almost exclusively in simulations that include collisional fragmentation (all points except the dark blue circles and black squares in the top panel).  This is consistent with our findings in \citetalias{clement18_frag}.  In that study, we concluded that added dynamical friction from ejected fragments tends to damp the eccentricities and inclination of growing Earth and Venus analogs \citep[see also:][]{chambers13,nader22}.
	\item \textbf{\textit{No satisfactory results were obtained in \citetalias{clement21_instb} instability models where Jupiter and Saturn inhabit a primordial 2:1 MMR.}}  As discussed in $\S$ \ref{sect:merc}, our \citetalias{clement21_instb}-style instabilities tend to overly-excite embryos in the Venus-forming region.  This is the result of the $\nu_{5}$ resonance being positioned in between Earth and Venus' modern orbits around time zero, and being stronger than in our \citetalias{nesvorny12} instabilities as a result of the gas giants' primordial eccentricity excitation.  This is clearly evidenced by the difference between the 2:1 (pink/red points in the bottom panel of figure \ref{fig:ei_ev}) and 3:2 (blue points) $\langle e,i \rangle _{EV}$ and $\Delta a_{EV}$ values, and the success rates for these metrics provided in table \ref{table:results}.
\end{itemize}

\subsection{Mars and the Asteroid Belt}
\label{sect:mars}

Our previous studies in \citetalias{clement18}, \citetalias{clement18_frag} and \citetalias{clement21_tp} were heavily focused on the ability of our models to replicate Mars' mass and rapid inferred accretion timescale \citep[e.g.:][]{Dauphas11}.  The success rates for $\tau_{Ma}$ and $M_{Ma}/M_{E}$ reported in table \ref{table:results} for our new simulations largely confirm our overall conclusions from those previous papers regarding the viability of the early instability scenario in terms of the Mars constraints.  While the primary focus of our current work is Mercury's formation, in this section we briefly build on the analyses of the early instability's effects on Mars and the asteroid belt presented our previous studies.

\subsubsection{Systems forming no Mars analog}

In \citetalias{clement21_tp} we noted that our new, GPU-grown \citetalias{clement20_psj} disks had slightly higher rates of forming no Mars analog when compared to the \citetalias{chambers98} disks considered in \citetalias{clement18} and \citetalias{clement18_frag}.  This is the result of the total planetesimal masses in the Mars-region being smaller in these models.  With fewer planetesimals available to damp the excited orbits of would-be Mars-analogs after the instability, the chances of losing all material from the region are higher.  Our new models confirm this finding, and we note that the effect is particularly pronounced in our \citetalias{clement21_instb} instability models that place Jupiter and Saturn in a 2:1 resonance with moderately enhanced eccentricities.  Specifically, 41$\%$ of our \citetalias{clement20_psj}$+$\citetalias{clement21_merc2}/\citetalias{clement21_instb} simulations form no Mars.  By comparison, our runs utilizing the identical terrestrial disk configuration in combination with the \citetalias{nesvorny12} instability model completely deplete the Mars-region only 13$\%$ of the time.  When combined with the other shortcomings identified in the previous sections, this does speak against the viability of the \citetalias{clement21_instb} model within the early instability framework.  However, we refrain from ruling the model out entirely as a reasonable fraction of our 2:1 instability models do produce adequate Mars analogs (table \ref{table:results}).  Interestingly, our \citetalias{hansen09} annulus instability models yield the lowest fraction of systems with no Mars analog (5$\%$)

\subsubsection{Consequences of the 2:1 Jupiter-Saturn resonance in the asteroid belt}

\begin{figure}
	\centering
	\includegraphics[width=.5\textwidth]{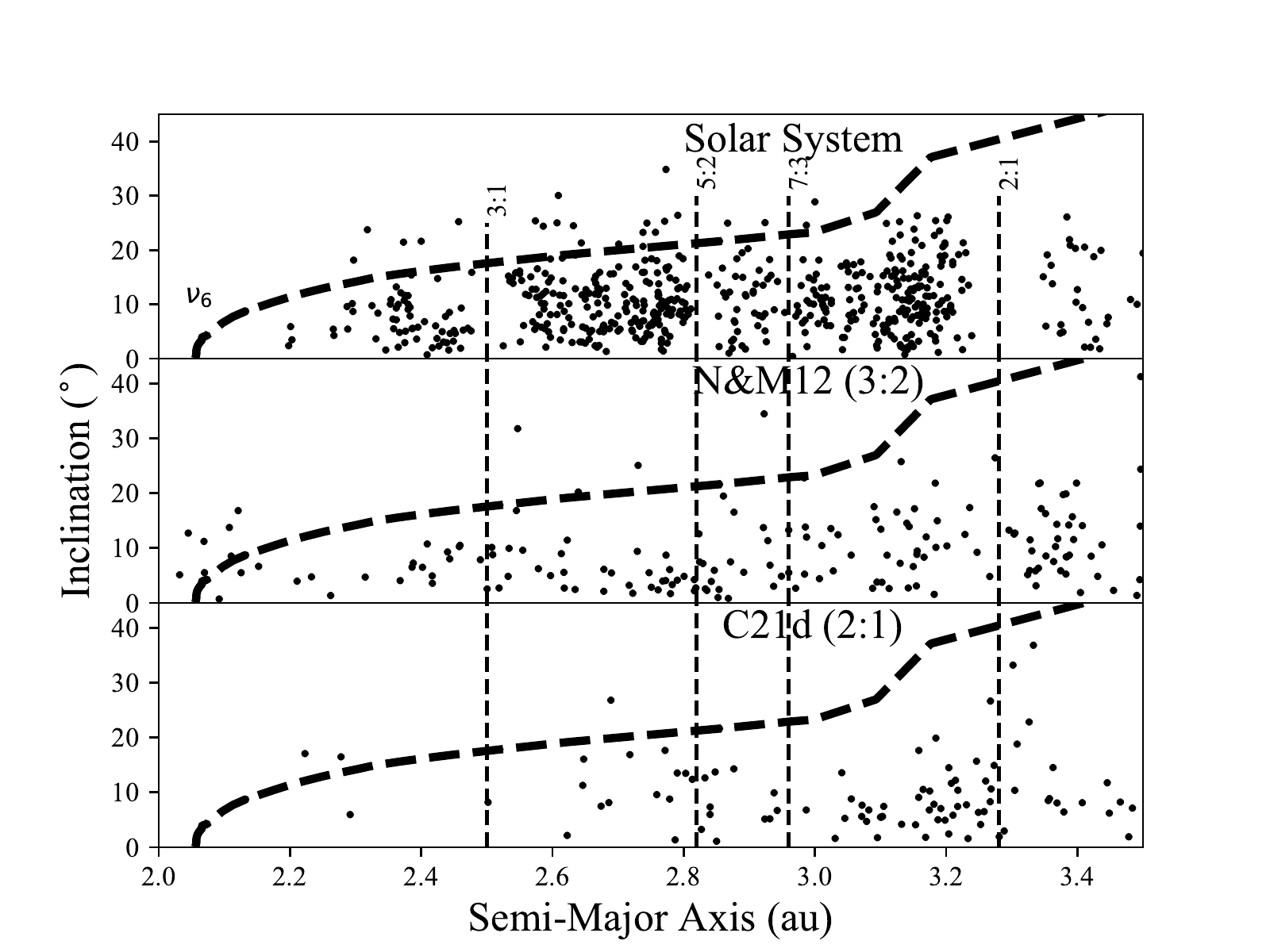}
	\caption{Orbital inclination versus semi-major axis distribution of large objects ($D\gtrsim$ 50 km) in the modern asteroid belt (top panel), compared with all surviving asteroids in our new simulations utilizing \citetalias{nesvorny12} instability models (middle panel) and \citetalias{clement21_instb} models (bottom panel).  The dashed lines annotate the locations of dominant mean motion and secular resonances in the belt.}
	\label{fig:ab}1
\end{figure}

Both our \citetalias{nesvorny12} and \citetalias{clement21_instb} instabilities are quite successful at heavily depleting a primordially massive asteroid belt.  The success of our instabilities initializing Jupiter and Saturn in the 2:1 resonance is potentially related to the fact that the giant planets' orbits are eccentric starting at time zero.  In \citetalias{clement21_tp} we did not analyze whether or not the 2:1 Jupiter-Saturn resonance negatively affects the asteroid belt's orbital structure.  Figure \ref{fig:ab} compares the orbital inclination distribution of large asteroids in the actual belt (top panel), with the co-added distributions of surviving asteroids in all of our new \citetalias{nesvorny12} (middle panel) and \citetalias{clement21_instb} (bottom panel) simulations.  With the exception of high-inclination asteroids in the inner belt that are expected to be removed during the giant planets' residual migration phase \citep{clement20_mnras}, both modeled distributions provide reasonable matches to the real belt.  It is noteworthy that the \citetalias{nesvorny12} 3:2 instabilities produce a better match to the inclination distribution of asteroids with $a<$ 2.5 au.  Indeed, the ratio of asteroids above, to those below the $\nu_{6}$ secular resonance (0.48; a success criteria we used in \citetalias{clement18} and \citetalias{clement18_frag}) is reasonably close to that of the actual asteroid belt (0.08).  However, the fact that our sample size of surviving asteroids is relatively small makes it difficult to draw any significant conclusions from this plot.

To first order our work confirms the compatibility of the \citetalias{clement21_instb} model with the asteroid belt's orbital structure.  We also compared the eccentricity distributions, as well as the relative populations of objects in different radial bins, and found them both to reasonably resemble the belt's modern structure.  However, co-adding the results of multiple simulations obviously introduces additional uncertainties to this analysis, and thus future work incorporating higher particle resolution \citep[e.g.:][]{deienno16,deienno18,clement18_ab} will be required to fully validate these conclusions.


\section{Discussion and Conclusions}

In \citetalias{clement18}, \citetalias{clement18_frag} and \citetalias{clement21_tp} we argued that an unusually early Nice Model \citep{Tsi05,levison08,nesvorny12,deienno17} instability occurring within the first 1-10 Myr after gas disk dissipation is capable of reducing the mass and formation timescale of Mars analogs \citep[a problem in classic terrestrial accretion models:][]{wetherill80_rev,chambers98,chambers01,ray09a}.  While such a timing for the giant planet instability is rather early when compared with other sequences of events considered in the literature \citep[e.g.:][]{nesvorny21_tp}, a myriad of recently derived constraints restrict the epoch of giant planet migration to the first $\sim$100 Myr after the solar system's birth \citep[see $\S$ \ref{sect:intro} and][for a recent review]{ray_morby_rev21}.  However, the physical process of nebular photo-evaporation might necessitate the solar system's instability occur at an epoch as early as considered in our past studies \citep{liu22}.  In this paper, we revisited the early instability scenario with new simulations aimed at modeling the planet Mercury's formation by incorporating a shorter integration timestep and additional planet-forming material interior to Venus' modern orbit.  While the initial conditions used in our simulations are broadly consistent with the outcome of models investigating the effects of an outward migrating proto-Venus and proto-Earth during the gas disk phase \citep{clement21_merc4}, in this paper we simply combine outer terrestrial disks that were tested against a number of constraints related to the outer three terrestrial planets in previous work \citepalias{clement21_tp} with inner disks that boosted the probability of forming adequate Mercury analogs as determined in a pair of recent papers \citepalias{clement21_merc3,clement21_merc2}.  This is an obvious weakness of our current modeling work, and we plan to test more realistic embryo and planetesimal distributions derived directly from a single gas-disk simulation in future work.

The main conclusion of our modeling efforts is that the early instability framework is consistent with all potential pathways for Mercury's formation considered here.  The majority of our analyses focus on differences in the distribution of simulation outcomes for systems initialized with various disk structures in both the Mercury-region and outer terrestrial disk, as well as two different initial giant planet configurations.  The primary results of varying each of these parameters are summarized below.

\subsection{The Mercury-forming region}
Our analysis of Mercury's formation focused on three different models developed in past work: (1) the terrestrial planets form from a narrow annulus of material from which Mercury is scattered out of and forms as a stranded embryo \citepalias{hansen09}, (2) a collection of 4-6 $\sim$Mars-mass planets form interior to Venus in a quasi-stable orbital configuration that is destabilized in a manner that yields many high-velocity, mantle-stripping collisions and leaves behind Mercury as the sole survivor \citepalias{clement21_merc3}, and (3) Mercury forms directly within a mass-depleted distribution of embryos and planetesimals in the inner solar system \citepalias{clement21_merc2}.  While each model produces at least one outstanding Mercury-Venus system, our best results occur in the \citetalias{clement21_merc3} lone survivor models.  In successful simulations, instability-induced eccentricity excitation enhances the rate of high-velocity collisions among the embryos, and enhances the rate at which all proto-planets in the region merge with Venus.  This process reduces the probability of forming Mercury with to large of a $P_{V}/P_{Me}$.  Contrarily, our \citetalias{hansen09} annulus models almost always yield $P_{V}/P_{Me}$ values that are too small, and the period ratios produced in the \citetalias{clement21_merc2} models are systematically too large.  Moreover, our \citetalias{hansen09} disks do not produce the high-speed collisions necessary for sufficiently altering Mercury's CMF, while our early instability models more regularly match this constraint.

\subsection{The outer terrestrial disk}

Our collection of early instability simulations initially partition embryos and planetesimals in the Venus, Earth and Mars-regions in three different manners: (1) a concentrated annulus of equal-mass embryos between 0.7-1.0 au \citepalias{hansen09}, (2) an extended disk of 100 embryos and 1,000 planetesimals with $M_{Emb}/M_{Pln}=$ 1.0 at all radial locations \citepalias{chambers98}, and (3) consistent with the outputs of GPU simulations of planetesimal accretion in a decaying gas disk \citepalias{clement20_psj}.  While the majority of our analysis focused on the generation of Earth and Venus' orbital spacing and degree of dynamical excitation, we also compared the potential Moon-forming impacts on Earth with the results of SPH modeling of the event \citep{canup04,canup12,reufer12,cuk12}.  In general, we find that both the \citetalias{hansen09} and \citetalias{clement20_psj} disks are capable of replicating the authentic Earth-Venus system after being evolved through the Nice Model instability (only one marginally successful result was obtained with the \citetalias{chambers98} disk).  While the \citetalias{hansen09} models produce the best statistical results, the solar system result lies near the low excitation and low orbital spacing extreme of the distribution of results for all modeled cases that include an instability.  Additionally, it is important to note that Earth's accretion is often too rapid in the annulus models \citep[e.g.:][]{kleine09}.  While we observe Moon-forming impact geometries that are analogous to those preferred in a range of studies in the literature, the most common events most closely resemble the canonical, low-velocity Mars-mass impactor model \citep{benz86,canup01,canup04b,canup04}.

\subsection{Instability model: 3:2 versus 2:1 Jupiter-Saturn resonance}

Similar to our methodology in \citetalias{clement21_tp}, our new simulations considered two giant planet instability models: (1) a model where Jupiter and Saturn emerge from the gas disk locked in a 3:2 MMR on circular orbits \citepalias{nesvorny12}, and (2) a second set of computations where the gas giants are captured in a 2:1 resonance with elevated eccentricities \citepalias{clement21_instb}.  While the differences between the two models are negligible in most of our analyses (e.g.: the asteroid belt's orbital structure, Mercury's CMF, and Mars' mass and formation time), we prefer the \citetalias{nesvorny12} model for two main reasons.  First, the strong resonant sweeping in the Mercury-forming region that is characteristic of our \citetalias{clement21_instb} simulations tends to entirely deplete the region of material.  Thus, the rate of forming no Mercury analog is much higher with the 2:1 Juptier-Saturn resonance.  This is also the case for the rate of forming no Mars analog.  Second, these enhanced perturbations also have the effect of over-exciting the Earth-Venus system.  As a result, the \citetalias{clement21_instb} Earth-Venus orbital spacings and levels of dynamical excitation are systematically larger than those of both the \citetalias{nesvorny12} instability models, and the actual solar system.

In conclusion, an early (t $\simeq$ 1-10 Myr after gas dissipation) Nice Model instability provides a compelling explanation for the dynamical configuration of all planets in the solar system.  While such a scenario is not the only plausible explanation for many of these qualities \citep[e.g.:][]{walsh11,bromley17,broz21,johansen21,izidoro22_ss}, all models must account for the giant planets' acquisition of their modern orbits.  Moreover, as demonstrated here, incorporating Mercury's formation in these types of models can yield additional insights and constraints.  Future studies in the mold of our current work should strive to increasingly incorporate novel cosmochemical, geophysical and dynamical constraints to break degeneracies between different formation scenarios and further pin down the authentic structure of the terrestrial planet forming disk in the solar system.

\section*{Acknowledgments}
The author's are grateful for the contributions of two anonymous reviewers.  N.A.K. thanks the National Science Foundation for support under award AST-1615975 and NSF CAREER award 1846388.  SNR is grateful to the CNRS's {\em Programme Nationale de Plantologie (PNP)}.  This work used the Extreme Science and Engineering Discovery Environment (XSEDE), which is supported by National Science Foundation grant number ACI-1548562. Specifically, it used the Bridges2 system, which is supported by NSF award number ACI-1445606, at the Pittsburgh Supercomputing Center \citep[PSC:][]{xsede}.
\bibliographystyle{apj}
\newcommand{\psj}{$PSJ$}
\newcommand{\sci}{$Science$ }
\bibliography{merc5.bib}

\end{document}